\documentclass[aps,prb,amsfonts,amsmath,amssymb,twocolumn,superscriptaddress,floatfix]{revtex4-2}

\usepackage{bm}
\usepackage[table,x11names]{xcolor}
\usepackage[breaklinks=true,unicode=true,urlcolor=RoyalBlue4,colorlinks=true, citecolor=RoyalBlue4,linkcolor=RoyalBlue4]{hyperref}
\hypersetup{pdftitle={Antlerite},pdfauthor={Anton},pdfsubject={Frustrated magnetism},pdfdisplaydoctitle}
\usepackage{graphicx}
\usepackage{booktabs}
\usepackage{multirow}
\usepackage{dcolumn}
\renewcommand{\vec}[1]{\bm{#1}}

\def\antH{Cu$_3$SO$_4$(OH)$_4$}
\def\antD{Cu$_3$SO$_4$(OD)$_4$}

\begin{document}

\title{Coupled frustrated ferromagnetic and antiferromagnetic quantum spin chains in the quasi-one-dimensional mineral antlerite, \antH}

\author{Anton A.\ Kulbakov}
\affiliation{Institut f\"ur Festk\"orper- und Materialphysik, Technische Universit\"at Dresden, 01069 Dresden, Germany}

\author{Denys Y.\ Kononenko}
\affiliation{Leibniz Institute for Solid State and Materials Research Dresden, 01069 Dresden, Germany}
\author{Satoshi Nishimoto}
\affiliation{Leibniz Institute for Solid State and Materials Research Dresden, 01069 Dresden, Germany}
\affiliation{Institut f\"ur Theoretische Physik, Technische Universit\"at Dresden, 01069 Dresden, Germany}

\author{Quirin Stahl}
\author{Aswathi Mannathanath Chakkingal}
\affiliation{Institut f\"ur Festk\"orper- und Materialphysik, Technische Universit\"at Dresden, 01069 Dresden, Germany}

\author{Manuel Feig}
\author{Roman Gumeniuk}
\affiliation{Institut f\"ur Experimentelle Physik, TU Bergakademie Freiberg, 09596 Freiberg, Germany}

\author{Yurii Skourski}
\author{Lakshmi Bhaskaran}
\author{Sergei A.\ Zvyagin}
\affiliation{Dresden High Magnetic Field Laboratory (HLD-EMFL), Helmholtz-Zentrum Dresden-Rossendorf, 01328 Dresden, Germany}

\author{Jan Peter Embs}
\affiliation{Laboratory for Neutron Scattering and Imaging, Paul Scherrer Institut, 5232 Villigen, Switzerland}

\author{In\'es Puente-Orench}
\affiliation{Instituto de Nanociencia y Materiales de Arag{\'o}n (INMA), CSIC-Universidad de Zaragoza, Zaragoza 50009, Spain}
\affiliation{Institut Laue-Langevin, 71 Avenue des Martyrs, CS 20156, CEDEX 9, 38042 Grenoble, France}
\author{Andrew Wildes}
\affiliation{Institut Laue-Langevin, 71 Avenue des Martyrs, CS 20156, CEDEX 9, 38042 Grenoble, France}

\author{Jochen Geck}
\affiliation{Institut f\"ur Festk\"orper- und Materialphysik, Technische Universit\"at Dresden, 01069 Dresden, Germany}
\affiliation{W\"urzburg-Dresden Cluster of Excellence on Complexity and Topology in Quantum Matter -- ct.qmat, Technische Universit\"at Dresden, 01069 Dresden, Germany}

\author{Oleg Janson}
\email{o.janson@ifw-dresden.de}
\affiliation{Leibniz Institute for Solid State and Materials Research Dresden, 01069 Dresden, Germany}

\author{Dmytro S.\ Inosov}
\email{dmytro.inosov@tu-dresden.de}
\affiliation{Institut f\"ur Festk\"orper- und Materialphysik, Technische Universit\"at Dresden, 01069 Dresden, Germany}
\affiliation{W\"urzburg-Dresden Cluster of Excellence on Complexity and Topology in Quantum Matter -- ct.qmat, Technische Universit\"at Dresden, 01069 Dresden, Germany}

\author{Darren C.\ Peets}
\email{darren.peets@tu-dresden.de}
\affiliation{Institut f\"ur Festk\"orper- und Materialphysik, Technische Universit\"at Dresden, 01069 Dresden, Germany}

\begin{abstract}

Magnetic frustration, the competition among exchange interactions, often leads to novel magnetic ground states with unique physical properties which can hinge on details of interactions that are otherwise difficult to observe.  Such states are particularly interesting when it is possible to tune the balance among the interactions to access multiple types of magnetic order.
  We present antlerite, \antH, as a potential platform for tuning frustration.  Contrary to previous reports, the low-temperature magnetic state of its three-leg zigzag ladders is a quasi-one-dimensional analog of the magnetic state recently proposed to exhibit spinon-magnon mixing in botallackite.  Density functional theory calculations indicate that antlerite's magnetic ground state is exquisitely sensitive to fine details of the atomic positions, with each chain independently on the cusp of a phase transition, indicating an excellent potential for tunability.

\end{abstract}

\maketitle

Magnetic frustration, wherein the exchange energy cannot be simultaneously minimized on all individual bonds in the spin system, leads to a wide array of novel magnetic phases\,\cite{Starykh2015,Diep2013,Kaplan2007}.  The ground state can be selected by a delicate balance of interactions, while the cancellation of stronger interactions can bring weaker interactions to the fore, allowing the observation of effects that would ordinarily be hidden or negligible\,\cite{Lacroix2011}.  Magnetic frustration can be achieved either geometrically, where spins populate a lattice whose spatial arrangement forces the interactions to compete, or through longer-range interactions which compete with shorter-range interactions\,\cite{Ramirez1994,Batista2016,Schmidt2017}.  A reduction in dimensionality can also assist in destabilizing conventional magnetic order, by reducing the number of exchange pathways that lower its energy.  The richest physics is expected where the energy scales of the interactions and the competition among them either prevent the system from finding a unique ground state, or make multiple spin configurations nearly degenerate, such that the material may be readily tuned among several exotic states.  This can be particularly interesting in quantum spin systems, materials with effective spin-$\frac12$ moments, where quantum fluctuations also play a significant role.

\begin{figure*}[ht]
\includegraphics[width=\textwidth]{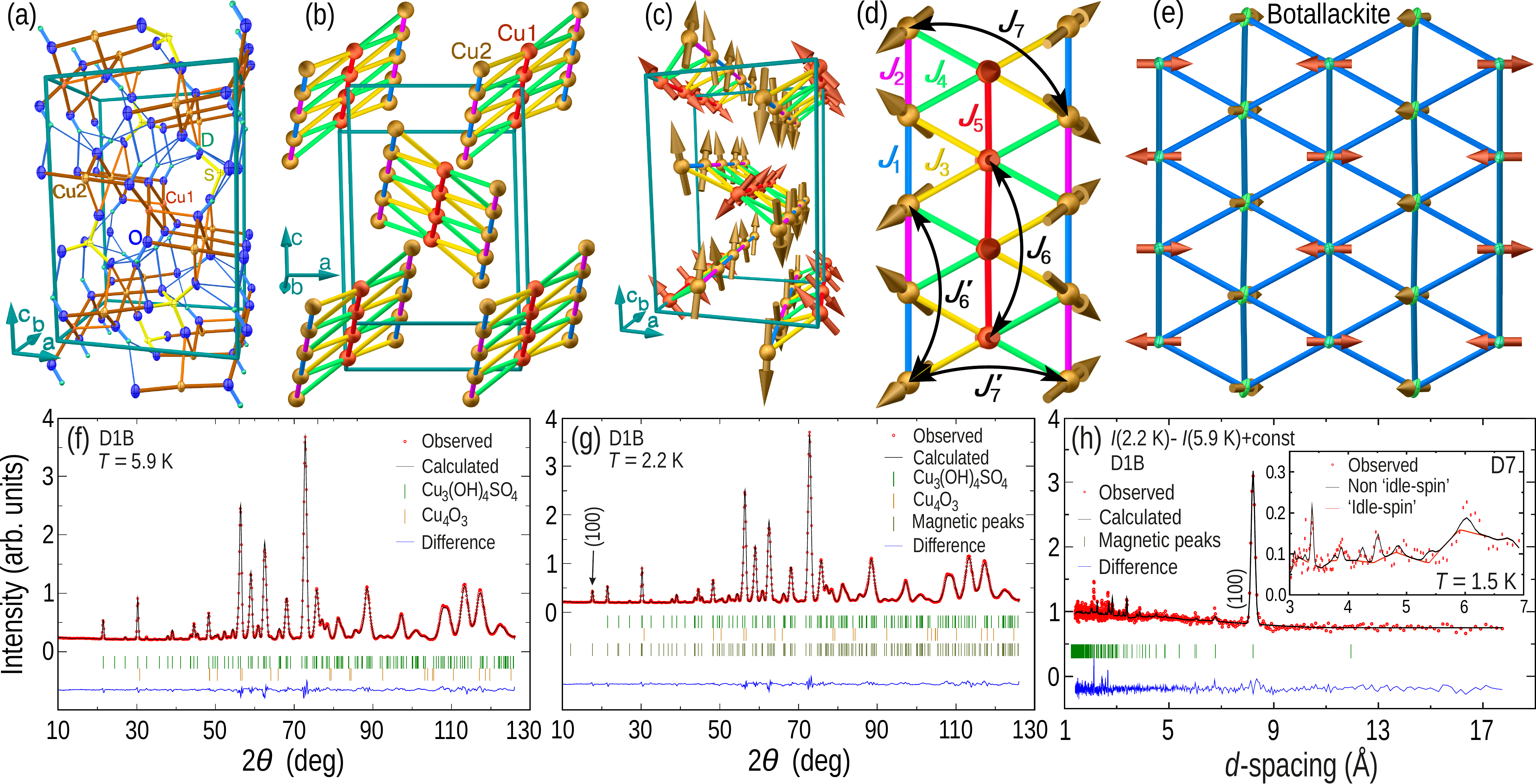}
\caption{(a) Crystal structure of \antD\ from single-crystal x-ray diffraction at 30\,K and 5.9-K neutron powder diffraction. (b) Cu sublattice viewed along the $\vec{b}$ axis. (c) Refined magnetic ground state of antlerite. (d) Exchange interactions within a ladder. (e) Cu sublattice\,\cite{Hawthorne1985} and spin orientations\,\cite{Zhang2020bot} in botallackite for comparison. Neutron powder refinements are shown for (f) the paramagnetic state at 5.9\,K and (g) the low-temperature ground state at 2.2\,K. (h) The difference between these datasets.  Inset: refinement of spin-polarized neutron diffraction data (D7) with the idle-spin and non-idle-spin magnetic models.}
\label{fig:fig1} 
\end{figure*}

Divalent Cu materials offer a particularly attractive playground for frustration, since a strong tendency toward Jahn-Teller distortions breaks orbital degeneracy, leading to a half-filled band, in which strong on-site interactions drive localization and favor $S$=$\frac12$ antiferromagnetism.  A wide variety of Cu sublattices are realized in natural minerals, predominantly composed of distorted Cu triangular motifs\,\cite{Inosov2018}, offering a rich playground for frustrated quantum spin physics.  
As two very recent examples, the coexistence of triplons, spinons, and magnons was reported in SeCuO$_3$\,\cite{Testa2021}, and the interaction of spinons and magnons was reported for the first time in botallackite Cu$_2$(OH)$_3$Br\,\cite{Zhang2020bot}.  In the latter, a distorted-triangular Cu plane can be understood as alternating ferro- and antiferromagnetic (AFM) 1D chains with weaker interchain interactions.
The magnetic ground state in botallackite, shown in Fig.~\ref{fig:fig1}(e), has now been explained through a combination of first-principles calculations based on density functional theory (DFT), linear spin-wave theory, and exact diagonalization\,\cite{Gautreau2021}, and related systems are now being explored theoretically as the spin model is generalized\,\cite{Majumdar2021}.  

The natural mineral antlerite, \antH, is a three-leg ladder compound in which zigzag bonds between the central and side legs form triangles of Cu$^{2+}$ ions\,\cite{Hawthorne1989} --- its copper sublattice is depicted in Fig.~\ref{fig:fig1}(b).  Such triangular-lattice ladders have been studied far less than their square-lattice analogues\,\cite{Smaalen1999},
despite the opportunity for strong frustration.  Previous neutron diffraction studies on antlerite indicated that only the outer legs of the ladder possess an ordered moment while the central leg exhibits unexplained ``idle-spin'' behavior\,\cite{Vilminot2003,Hara2011}.  However, follow-up studies have seriously questioned this result.  DFT calculations showed strong AFM coupling along the central leg of the ladder, which would be expected to induce order\,\cite{Koo2012}.
Then, comprehensive specific-heat and proton-NMR measurements in magnetic field\,\cite{Fujii2013} found a more complex magnetic phase diagram than previously reported\,\cite{Hara2011}, including a phase which can only be realized if the system has at least four distinct magnetic sites.  The correct low-temperature magnetic ground state has remained elusive.  

In this Letter, we determine the low-temperature magnetic ground state of antlerite and model its magnetic interactions using DFT. It is not idle spin as previously proposed\,\cite{Vilminot2003}.  In a ground state strongly reminiscent of botallackite, the ferromagnetic (FM) outer legs of the ladder are antialigned but noncollinear, while the central leg is AFM with a very different spin orientation.  DFT finds that both inner and outer chains are on the cusp of phase transitions.  This proximity to multiple quantum phase transitions, and the complex phase diagrams already reported\,\cite{Hara2011,Fujii2013}, suggest a unique ability to tune a state of coupled FM and AFM chains.  We anticipate that antlerite could serve as a versatile platform for investigating spinon-magnon interactions and the competition among magnetic ground states.

Antlerite was synthesized hydrothermally under autogenous pressure at 180\,$^\circ$C in a Teflon-lined stainless steel autoclave, from CuSO$_4\cdot 5$H$_2$O (Alfa Aesar, 99\,\%) and Cu(OH)$_2$ (Alfa Aesar, 94\,\%) in distilled or deuterated (Acros Organics, 99.8\,\%D) water.  Single-crystal x-ray diffraction data were collected from 30 to 295\,K on a Bruker-AXS KAPPA APEX II CCD diffractometer with graphite-monochromated Mo-\(K_{\alpha}\) radiation using an Oxford Cryosystems N-HeliX cryostat to verify the space group symmetry and atomic positions. Weighted full-matrix least-squares refinements on $F^2$ were performed with {\sc Shelx}\,\cite{Shelx2008,Shelx2015} as implemented in {\sc WinGx} 2014.1\,\cite{WinGX}. To determine accurate hydrogen positions and the magnetic structure, high-resolution neutron powder diffraction was performed at D1B\,\cite{D1B} and neutron diffraction with polarization analysis using D7\,\cite{D7}, both at the Insitut Laue-Langevin (ILL), Grenoble, France. Powder diffraction data were Rietveld-refined in {\sc FullProf} by the full-matrix least-squares method\,\cite{FullProf}, using the scattering factors from Ref.~\onlinecite{Sears1992}.  The atomic positions, shown in Fig.~\ref{fig:fig1}(a) and detailed in the Supplemental Material in Section~\ref{suppXRD}, largely agree with the previously reported structure\,\cite{Hawthorne1989,Vilminot2003}, hydrogen positions excepted.  The latter are now better aligned with their host oxygen atoms along $c$.

Powder inelastic neutron scattering (INS) was measured on the FOCUS spectrometer at the Paul Scherrer Institute, Villigen, Switzerland\,\cite{FOCUS1997}, using wavelengths of 5 and 6\,\AA, having energy resolution of 91 and 43\,$\mu$eV at the elastic line, respectively, and was modelled using {\sc SpinW}\cite{SpinW}. High-field magnetization of a powder sample was measured at the Dresden High Magnetic Field Laboratory (HLD), Helmholtz-Zentrum Dresden-Rossendorf (HZDR), Germany, using a 60-T pulsed field magnet with a rise time of 7\,ms and a total pulse duration of 25\,ms.  The  magnetization  was obtained by integrating the voltage induced in a compensated coil system surrounding the sample \cite{magn}.

Neutron diffractograms in the paramagnetic state at 5.9\,K and in the low-temperature state at 2.2\,K are shown in Figs.~\ref{fig:fig1}(f) and \ref{fig:fig1}(g), respectively. Additional magnetic intensity on structural Bragg peaks and new weak magnetic peaks were observed at low temperature, as seen in the difference spectrum in Fig.~\ref{fig:fig1}(h) --- the strongest magnetic reflection, the structurally forbidden (100), is highlighted.  Our refinements of the low-temperature ground state do not support the idle-spin model proposed previously\,\cite{Vilminot2002,Vilminot2003,Hara2011}, which fails to predict several observed magnetic peaks as shown in the inset to Fig.~\ref{fig:fig1}(h).  The data are instead best described by a model with propagation vector $(100)$ having AFM order along the central (Cu1) leg of the ladder and significant canting of the other spins, as depicted in Fig.~\ref{fig:fig1}(c).  The ordered moments on the Cu1 site are 0.80(22)\,$\mu_\text{B}$ along ($\pm$0.54(17), 0, $\pm$0.59(14)), while the ordered Cu2 moments of 0.97(5)\,$\mu_\text{B}$ lie along (0, $\pm$0.38(5), $\pm$0.89(9)) and are ferromagnetically aligned along each side leg. As found previously, the side legs are mutually antialigned.
Accounting for $g$ factors of 2.1--2.3 from ESR\,\cite{Okubo2009}  (also see Supplemental Material in Section~\ref{suppESR}) and magnetization\,\cite{Hara2011}, this indicates that Cu2 is nearly fully ordered, while a reduced Cu1 moment may arise from the quantum fluctuations expected for $S=\frac12$, from frustration, or from the low number of interactions present in this low-dimensional material.  Our refined ground state resembles the ``AF6'' state calculated in Ref.~\cite{Koo2012} but with significant additional canting.

To determine the isotropic magnetic exchange interactions in antlerite, we performed density-functional-theory (DFT) band-structure
calculations using the full-potential local-orbital code {\sc FPLO} v.\ 18\,\cite{FPLO}. We employed a scalar-relativistic treatment within the
generalized gradient approximation (GGA) for the exchange and correlation
potential\,\cite{PBE96}. For the structural input, we used our newly refined low-temperature unit
cell parameters and atomic coordinates, which are described alongside further details of our calculations in the Supplemental Material below.

In a nonmagnetic GGA calculation, antlerite features a spurious metallic ground
state due to the underestimation of Coulomb repulsion among the Cu $3d$
electrons.  An eight-band manifold crossing the Fermi
level is due to the antibonding combination
of Cu~$3d_{x^2 - y^2}$ and O~$2p_{\sigma}$ states. We mapped these states onto
an effective one-orbital model which is parameterized by projecting the
respective GGA bands onto a Wannier basis of Cu $3d_{x^2 - y^2}$ states. 
The parameters of the effective one-orbital model are hopping integrals
$t_{ij}$ that describe virtual electron transfer between Cu sites $i$ and $j$.
Nine $t_{ij}$ terms exceeding the threshold of 15\,meV are provided in
Table~\ref{tab:js}, where we adopt the notation from Refs.~\onlinecite{Vilminot2003,
Hara2011,Koo2012}. In agreement with the previous
works, we find that antlerite is a quasi-one-dimensional (1D) magnet, with sizable
hopping interactions confined to the three-leg ladders\,\footnote{Interladder exchange interactions are on the order of fractions of a kelvin, two orders of magnitude weaker than the dominant intraladder exchanges.  While they play a key role in producing three-dimensional long-range order, they have negligible impact on the type of order or the physics within a ladder}. 

To estimate the respective exchange integrals, we doubled the unit cell in the
$\hat{\vec{b}}$ direction and constructed 64 inequivalent magnetic
configurations, whose total energies were calculated on a 2$\times$2$\times$2 $k$-mesh within the GGA+$U$ approximation. To describe
interaction effects in the $3d$ shell of Cu, we used the onsite Coulomb
repulsion $U$\,=\,8.5\,eV, the onsite Hund exchange $J$\,=\,1\,eV, and the
fully localized limit for the double counting correction.  The resulting total
energies are mapped onto a classical Heisenberg model, whose model parameters---the magnetic exchange integrals $J_{ij}$ in Table~\ref{tab:js}---are
evaluated by a least-squares solution (see Supplemental Material in Section~\ref{suppDFT}).

While both the crystal and electronic structure of antlerite are shaped by ladders,
the backbone of its spin model is the legs, which are coupled by relatively weak $J_3$ and $J_4$ exchanges.
The central leg hosts competing AFM exhanges $J_5$ and $J_6$ 
operating, respectively, between first and second neighbors. In contrast, the
side legs feature alternating first-neighbor FM exchanges
$J_1$ and $J_2$ and a weaker AFM exchange $J_6'$ between second
neighbors the relevance of which, to the best of our knowledge,
has not been discussed previously. Finally, while two relevant hoppings connect the
side legs, only $J_7$ gives rise to a small AFM exchange.
Our microscopic magnetic model is fundamentally different from the
phenomenological model of Ref.~\onlinecite{Vilminot2003} and qualitatively similar to
the band-structure-based model of Ref.~\onlinecite{Koo2012}. Despite this qualitative
similarity, the ratios such as $J_6/J_5$ and $J_3/J_5$, as well as the absolute
values of the exchanges, differ significantly, possibly because of the different choice of $U$, different code, or a high sensitivity of antlerite to perturbations.

\begin{table}[tb]
  \caption{\label{tab:js}Leading hopping ($t_{ij}$, in meV) and exchange
($J_{ij}$, in K) integrals in antlerite based on our refined structure. Positive exchanges are AFM. The respective intersite distances
$d_{\text{Cu--Cu}}$ are given in \AA.}
    \begin{tabular}{rrrrc@{~}l}\toprule\toprule
      Label & $d_{\text{Cu--Cu}}$ & $t_{ij}$ & $J_{ij}$ & & Type of exchange \\ \midrule
      $J_1$  & 2.980 &  93 &$-26$      & \multirow{2}{*}{$\Bigl\}$} & \multirow{2}{*}{first-neighbor in outer legs} \\
      $J_2$  & 3.053 &  73 &$-11$      & \\
      $J_3$  & 3.240 & 101 &   9       & \multirow{2}{*}{$\Bigl\}$} & \multirow{2}{*}{couple central and outer legs} \\
      $J_4$  & 3.151 & 103 &  11       & \\
      $J_5$  & 3.018 & 167 &  48       & & first-neighbor in central leg \\
      $J_6$  & 6.034 &  56 &  25       & & second-neighbor in central leg \\
      $J_6'$ & 6.034 &  18 &   6       & & second-neighbor in outer legs \\
      $J_7$  & 6.391 &  26 &   1       & \multirow{2}{*}{$\Bigl\}$} & \multirow{2}{*}{couple outer legs} \\ 
      $J_7'$ & 5.634 &  30 & $\simeq0$ & \\ \bottomrule\bottomrule
\end{tabular}
\end{table}

We start the discussion of the low-temperature magnetic ground state by considering a
simplified spin model of decoupled legs ($J_3=J_4=0$). Here, the central
leg would have a helical ground state in the classical model and a gapped phase in the
quantum $S=\frac12$ case. The side legs would form helices, but in the
vicinity of the fully polarized (ferromagnetic) phase; both states are
quasi-classical with minor quantum corrections.

When we reinstate the interleg exchanges $J_3$, $J_4$, and $J_7$, the
consequent leverage has a drastic impact on the magnetism. At the classical
level, a noncollinear state with the central leg twisted into a helix and fully polarized side legs has slightly lower energy than the collinear state which corresponds to the experimental (100) propagation vector.
This disagreement may arise from inaccuracies in the
exchange integrals, as a collinear state adiabatically connected to the (noncollinear) experimental ground state is found only $\sim$1.5\,K higher in
energy than the helical state, well within the uncertainty of DFT.  Such a small energy difference suggests an exquisite sensitivity to the ratios of the exchange interactions, which would imply that the competition between collinear and helical states can be
manipulated by perturbations such as chemical substitution, pressure, or magnetic field.

To determine the ground state of the quantum model, we performed density-matrix
renormalization group (DMRG) simulations\,\cite{White1992,White1993} using an open cluster of $40 \times 3$ spins; further details are in the Supplemental Material below. The resulting spin correlations indicate a noncollinear state in which
central as well as side legs are twisted into a helix. Similar to the
classical model, other states have comparable energies. For instance, the
enhancement of $|J_1|$ and $|J_2|$ and the reduction of $J_6'$ readily
stabilize the correct ground state~\footnote{The small mutual noncollinearity of side
legs stems from the exchange anisotropy and cannot be accounted for in the
(isotropic) Heisenberg model}. More details on the energy balance of the
competing phases are provided in the Supplemental Material below. 

\begin{figure}[htbp]
    \includegraphics[width=\columnwidth]{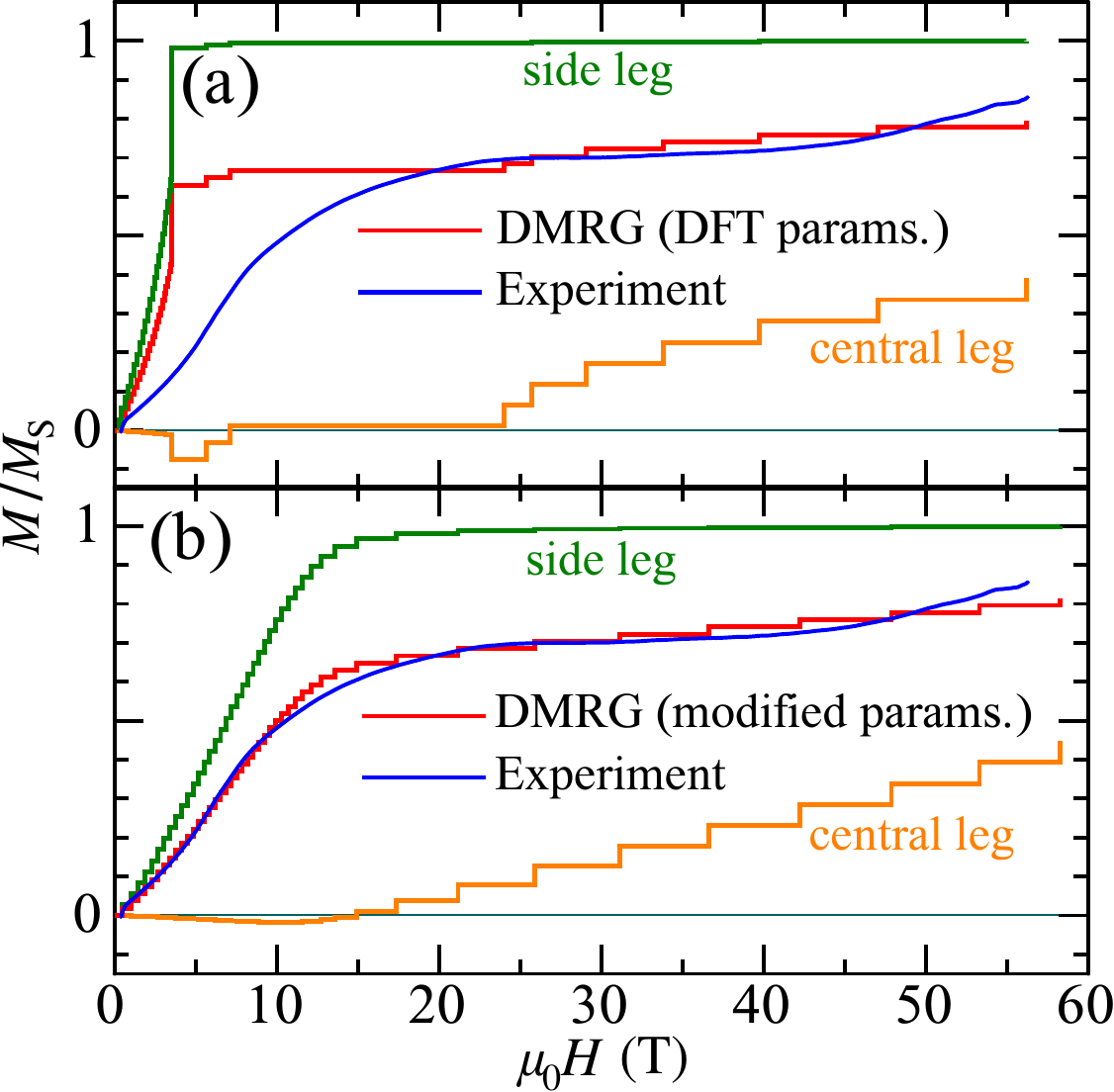}
    \caption{\label{fig:mh}Comparison of the magnetization curve
      measured on \antH\ powder at 1.4\,K (blue) against the DMRG simulation of the spin model (red). (a) Exchanges from Table~\ref{tab:js} based on our refined crystal structure result in abrupt polarization of the side legs. (b) A modified parameter set (see text) improves the agreement. A $g$ factor of 2.18 taken from ESR (Fig.~\ref{fig:ESR}) was used for scaling $B$.}
\end{figure}

To gain further confidence in the exchange parameters, we computed the
longitudinal magnetization $\langle S^z\rangle$ by DMRG on a $36 \times 3$
cluster where periodic boundary conditions were applied along the ladder, and compared it with the experimental magnetization curve. While
the field dependence of the magnetization for the parameter set from
Table~\ref{tab:js} is too steep [Fig.~\ref{fig:mh}(a)], an effective description based on a modified
set of parameters ($J_1=-25.2$, $J_2=-16.8$, $J_3=14.7$,
$J_4=6.3$, $J_5=42.0$, $J_6=10.5$, $J_6'=1.7$, and $J_7=6.7$\,K,
also see the Supplemental Material below) yields excellent agreement
[Fig.~\ref{fig:mh}(b)].  With the data at hand, we conclude that antlerite
embodies a delicate balance of frustrated interactions operating within the
central and side legs as well as connecting them into a ladder.  Furthermore,
by applying small perturbations, the
appearance of fascinating phenomena caused by competition/collaboration between the
Haldane physics and order-by-disorder mechanism is highly likely.

\begin{figure*}[htb]
\includegraphics[width=\textwidth]{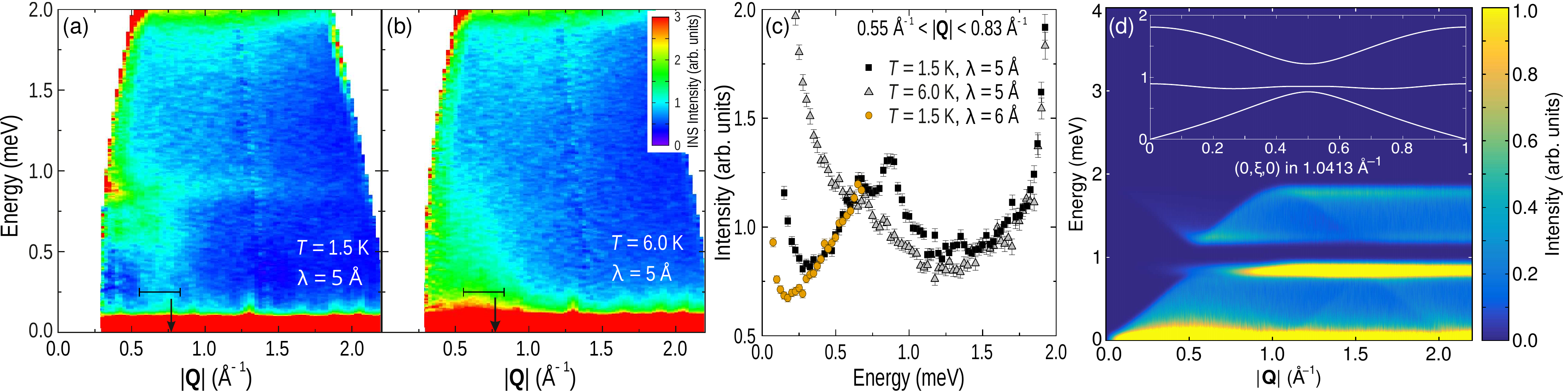}
\caption{\label{FOCUS}Powder INS spectra of deuterated antlerite.  Intensity maps at (a) 1.5 and (b) 6.0\,K.  (c) Temperature dependence of the intensity, averaged over the $Q$ range identified by bars in (a,b); the 6-\AA\ data were rescaled to match the 5-\AA\ data at high energies.  (d) Simulated spectra based on themodified DMRG parameters; main panel is powder-averaged.}
\end{figure*}

Finally, we comment on the magnetic state above the plateau-like feature at $M/M_\text{s}=2/3$ for saturation magnetization $M_\text{s}$, which is not a plateau so much as a gradual upturn. This means that quantum fluctuations are strong, and the central leg is no longer fully AFM ordered, likely due to a loss of stabilizing interactions with the now-polarized side legs, which would also act to screen it from the neighboring ladders. Since this would render the central leg as an essentially ideal 1D Heisenberg spin system and an exotic field-induced Tomonaga-Luttinger liquid (TLL), it is worth looking closely at the DMRG results in this regime.

\begin{figure}[b!]
  \includegraphics[width=\columnwidth]{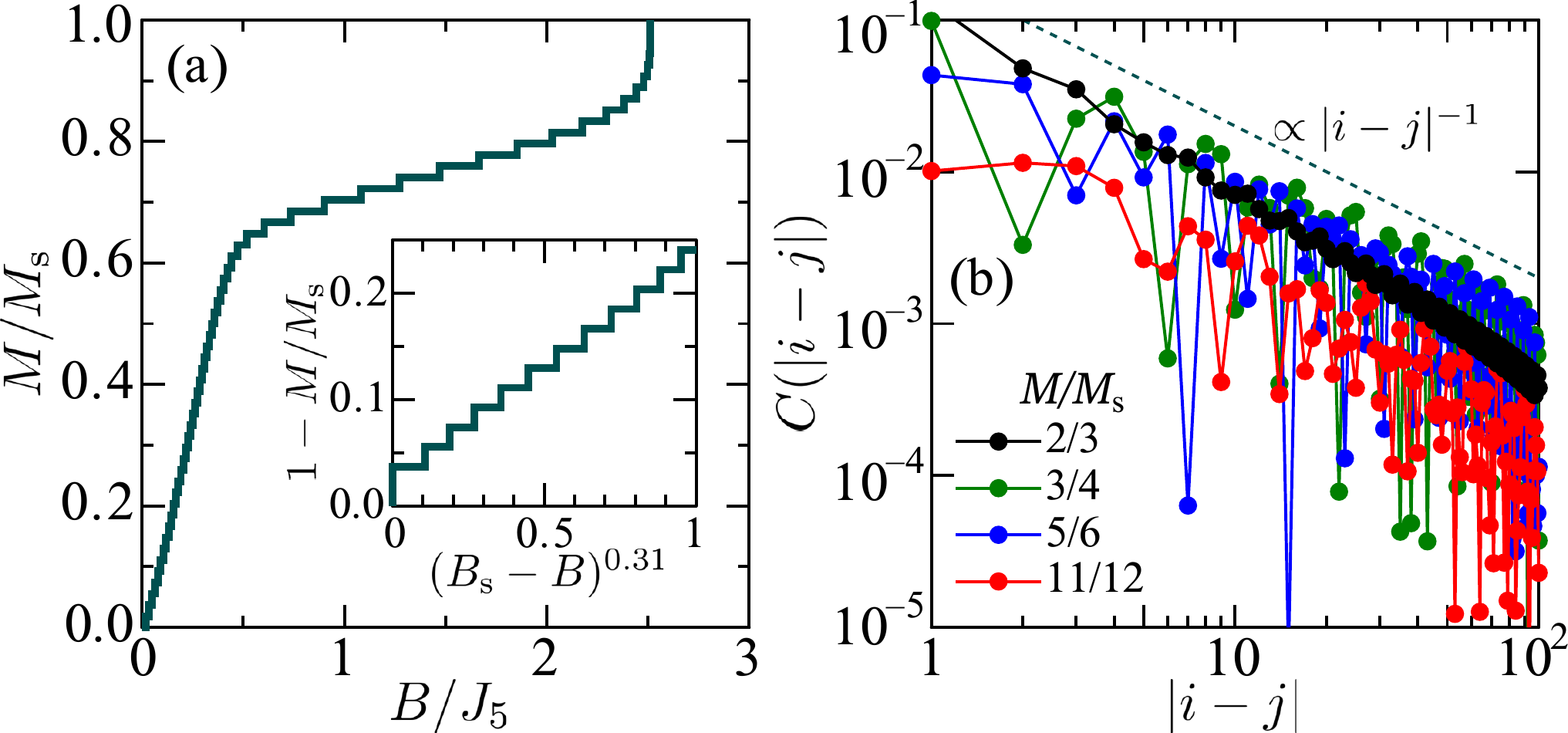}
    \caption{(a) Full magnetization curve calculated by DMRG with the modified parameter set. Inset: Singular behavior of $1-M/M_\text{s}$ {\itshape vs}.\ $(B-B_\text{s})^\gamma$ with $\gamma=0.31$ near the saturation field.  (b) Log-log plot of the spin-spin correlation function $C(|i-j|)$ as a function of distance along the central leg for several $M/M_\text{s}$ values.  The asymptotic behavior of the pure 1D AFM Heisenberg model, $\propto |i-j|^{-1}$, is shown as a guide. \label{fig:TLL}}
\end{figure}

We find that the central leg may be regarded as an ideal 1D Heisenberg spin system at $M/M_\text{s}\ge2/3$, implying a field-induced TLL phase. To confirm the TLL nature of the central leg at $M/M_\text{s}\ge2/3$, we investigated the shape of the full magnetization curve [plotted {\itshape vs}.\ $B/J_5$ in Fig.~\ref{fig:TLL}(a)] as well as the asymptotic behavior of the spin-spin correlation function.  At $M/M_\text{s}\sim2/3$, the side legs are completely polarized by the field while the spin state of the central leg remains nearly singlet. Upon increasing $B$, the magnetization of the central leg then exhibits an initially slow increase and ultimately a singularity $dM/dB=\infty$ at the saturation field $B_\text{s}/J_5\sim2.5$. This behavior is typical of a 1D AFM Heisenberg system in a critical regime. Using perturbation theory, the form of the magnetization curve as $M \to M_\text{s}$ is known to be $1-M/M_\text{s}\propto(B-B_\text{s})^\gamma$, with $\gamma>0$\,\cite{Parkinson1985}. Our calculated magnetization curve indeed takes this form, with $\gamma=0.31$, as shown in the inset of Fig.~\ref{fig:TLL}(a).  This $\gamma$ value deviates from that of the pure 1D AFM Heisenberg chain with only nearest-neighbor coupling ($\gamma=0.5$) due to the existence of next-nearest-neighbor coupling $J_6$.  The spin-spin correlation function $C(|i-j|)=\langle S^z_iS^z_j\rangle-\langle S^z_i \rangle\langle S^z_j \rangle$ calculated by DMRG with a $160\times3$ open cluster is plotted as a function of distance $|i-j|$ in Fig.~\ref{fig:TLL}(b) for several $M/M_\text{s}~(\ge2/3)$ values.  In every case, a power-law decay is clearly evident. Interestingly, the decay ratios seem to be relatively close to the $\propto |i-j|^{-1}$ of the pure 1D AFM Heisenberg chain despite $J_6$, confirming that the central leg is a field-induced TLL for $2/3 \le M/M_\text{s} \le 1$.  

Powder INS spectra collected at 1.5 and 6.0\,K to validate the model parameters are shown in Figs.~\ref{FOCUS}(a,b) and to higher energy in Fig.~\ref{moreFOCUS} in the Supplemental Material.  Peaks in the magnetic intensity at energies of $\sim$0.6 and 0.8\,meV may correlate to the narrow band and the bottom of the upper magnon band in Fig.~\ref{FOCUS}(d), calculated based on the DMRG exchange parameters above.  This intensity vanishes for higher momentum transfer $Q$ and higher temperatures [Fig.~\ref{FOCUS}(c)] as expected for a magnetic signal, and appears to disperse to a minimum near the propagation vector (black arrows).  The calculated exchange parameters can evidently describe the key features of the spectrum, aside from an overall scale factor that is likely associated with the uncertainties of DFT or quantum renormalization, supporting the ratios among our $J$ parameters.  Flat magnon bands have been predicted in sawtooth and kagome lattices\,\cite{Derzhko2015}, where they may be a potential platform for magnon crystals, nontrivial magnon topology, and non-Hermitian exceptional points\,\cite{McClarty2020}, and it would be interesting if the narrow feature around 0.8\,meV proved to be such a flat magnon band.

The low-temperature magnetic state we find in \antH\ contains all the essential ingredients found in botallackite: a low-dimensional Cu$^{2+}$ spin system with mutually antialigned FM legs, separated by an AFM leg with a spin orientation tilted by a significant angle relative to the FM spin orientation, linked through a distorted-triangular motif.  This suggests that antlerite may also realize the spinon-magnon mixing recently proposed in botallackite\,\cite{Zhang2020bot} and thus far reported in no other materials.  The key difference is that in botallackite these legs alternate in a two-dimensional sheet, whereas in antlerite they form quasi-one-dimensional three-leg ladders with very weak interladder coupling.  In antlerite, we find that the magnetic ground state is on the cusp of several other potential ground states, which are closely spaced in energy, suggesting that the magnetic ground state may be readily tuned, for instance by applied field, strain, or chemical substitution.  In fact, when we first calculated the magnetic ground state starting from the atomic positions reported in Ref.~\onlinecite{Vilminot2003}, we obtained a Haldane-like up-up-down-down magnetic ground state distinct from those discussed here.  Since the reported structure differs from ours chiefly in interladder hydrogen positions irrelevant to the intrachain interactions, this is further evidence of an extreme sensitivity to perturbations. Besides idle-spin cycloidal order reported upon isoelectronic substitution of S by Se\,\cite{Vilminot2007}, which could perhaps be revisited in light of the non-idle-spin magnetism found here, a cascade of phase transitions have been reported with temperature and in applied magnetic fields\,\cite{Hara2011,Fujii2013,Kulbakov2022b}, to which we now add the high-field plateau.  We also note that szenicsite Cu$_3$MoO$_4$(OH)$_4$ has a similar three-leg ladder motif\,\cite{Fujisawa2011}, but with a different alternation of exchange parameters which leads the outer chains to dimerize into a classical spin-1 system\,\cite{Lebernegg2017}.  The synthesis of szenicite has not been reported, but substitution of S by Mo may offer not only a further route for tuning the magnetism, but feedback on how to grow this Mo analog.

The distorted triangular lattice motifs found in Cu-based minerals have already revealed a number of interesting exotic magnetic phases, and antlerite stands to serve as an excellent platform for exploring their wealth of physics.  The interchain exchanges in \antH\ give rise to a model with a rich phase diagram that should be readily explored experimentally thanks to the advantageous energy scales.  

\begin{acknowledgments}
We thank Ulrike Nitzsche for technical assistance.
This project was funded by the German Research Foundation (DFG) via the projects A05, C01, C03, and C06 of the Collaborative Research Center SFB 1143 (project-id 247310070); GRK 1621 (project-id 129760637); the W\"urzburg-Dresden Cluster of Excellence on Complexity and Topology in Quantum Matter--{\slshape ct.qmat} (EXC~2147, project-id 390858490); through individual research grants IN~209/9-1 
and PE~3318/2-1
, and through project-id 422219907.  D.K.\ and O.J.\ were supported by the Leibniz Association through the Leibniz Competition.
The authors acknowledge the Institut Laue-Langevin, Grenoble (France) for providing neutron beam time\,\cite{ILL-data1_D1B-2020,ILL-data_D7-2021}, and the HLD at HZDR, member of the European Magnetic Field Laboratory (EMFL).  Part of this work is based on experiments performed at the Swiss spallation neutron source SINQ, Paul Scherrer Institute, Villigen, Switzerland.

\end{acknowledgments}

\bibliography{antlerite,DFG}

\clearpage
\renewcommand{\thefigure}{S\arabic{figure}}
\renewcommand{\thetable}{S\Roman{table}}
\renewcommand{\thesection}{S\arabic{section}}
\renewcommand{\theequation}{S\arabic{equation}}
\section{Supplemental Material\label{supp}}
This Supplemental Material provides further details on experimental procedures, structure refinements, density functional theory (DFT) calculations, and density-matrix renormalization group (DMRG) simulations. CIF files for our refined crystal and magnetic structures are provided as arXiv ancillary files at \href{https://arxiv.org/src/2203.15343v2/anc}{arxiv.org/src/2203.15343v2/anc}.

\section{Further experimental details}        

Neutron powder diffraction data were collected at D1B, ILL, Grenoble, France from 1 to 129$^\circ$ in steps of 0.1$^\circ$ using neutrons with a calibrated wavelength $\lambda = 2.528605$\,\AA, selected with a highly-oriented pyrolytic graphite [002] monochromator; instrumental broadening was determined from the refinement of a Na$_2$Ca$_3$Al$_2$F$_{14}$ standard, while wavelengths were refined using Si. Parasitic diffraction peaks from the sample environment were eliminated by a radial oscillating collimator. To more clearly resolve the difference between idle- and non-idle-spin ground states, additional neutron diffraction data with polarization analysis, shown in the inset to Fig.~1(h) in the main text, were collected on the D7 diffractometer at the ILL, Grenoble, France, at 1.5\,K using a neutron wavelength of 4.8707\,\AA.

Three-dimensional x-ray data were integrated and corrected for Lorentz, polarization and background effects using Bruker's {\sc Apex3} software\,\cite{Bruker}. Lattice parameters from our 5.9-K powder neutron diffraction data were used for the 30-K x-ray refinement due to higher reliability.  The x-ray atomic positions of all atoms other than hydrogen were then used as the basis for neutron powder refinements to determine the hydrogen positions, which were then introduced into the x-ray refinement as fixed parameters, in an iterative process. All neutron data were collected on deuterated samples ($\sim$95\,\% deuteration), to avoid the enormous incoherent cross-section of $^1$H.  X-ray techniques are sensitive to electron density, which in the case of hydrogen is minimal and shifted relative to the nucleus, so a direct comparison of H positions from x-ray diffraction with D positions from neutron diffraction would not be meaningful.

\begin{figure}[tbp]
  \includegraphics[width=\columnwidth]{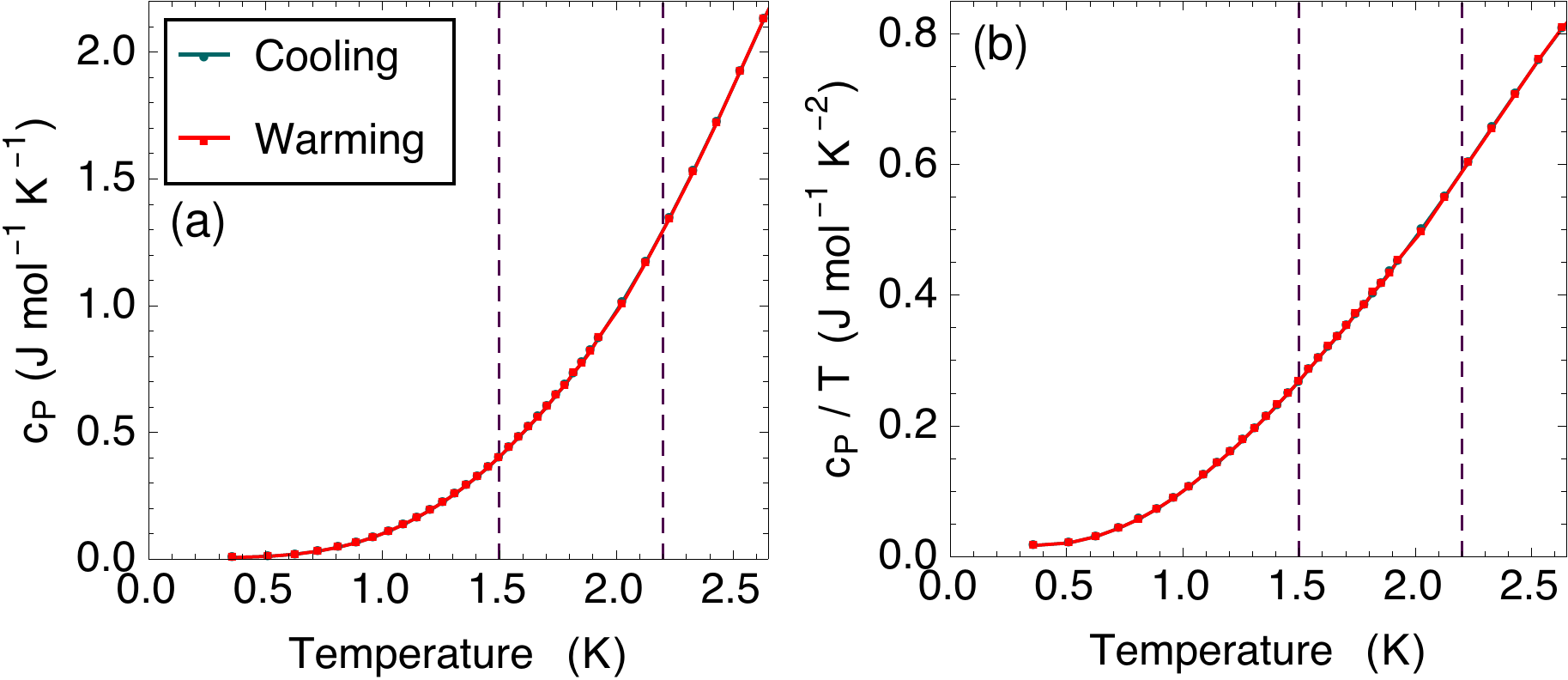}
  \caption{\label{fig:cP}(a) Low-temperature specific heat $c_P$ and (b) $c_P/T$ of \antH.  The dashed lines denote temperatures at which magnetic diffraction was performed.  No transitions are observed near or below these temperatures, indicating that we have probed the low-temperature magnetic ground state.}
\end{figure}

Magnetization data collected down to 1.8\,K on a Quantum Design Magnetic Properties Measurement System (MPMS3-VSM) magnetometer (not shown) and specific heat data measured down to 0.35\,K using the heat capacity and helium-3 options on a Quantum Design Physical Properties Measurement System (PPMS-DynaCool) are consistent with the phase diagram proposed in Ref.~\onlinecite{Fujii2013} insofar as there are no transitions observed below 2.8\,K in low field (see Fig.~\ref{fig:cP}).  These data give us confidence that our magnetic diffraction experiments at 1.5 and 2.2\,K were performed in the ground state phase.  The lack of an  electronic contribution to the specific heat at low temperatures, together with antlerite being transparent and green, lead us to conclude that it is an insulator.  

\begin{figure*}[htbp]
  \includegraphics[width=\textwidth]{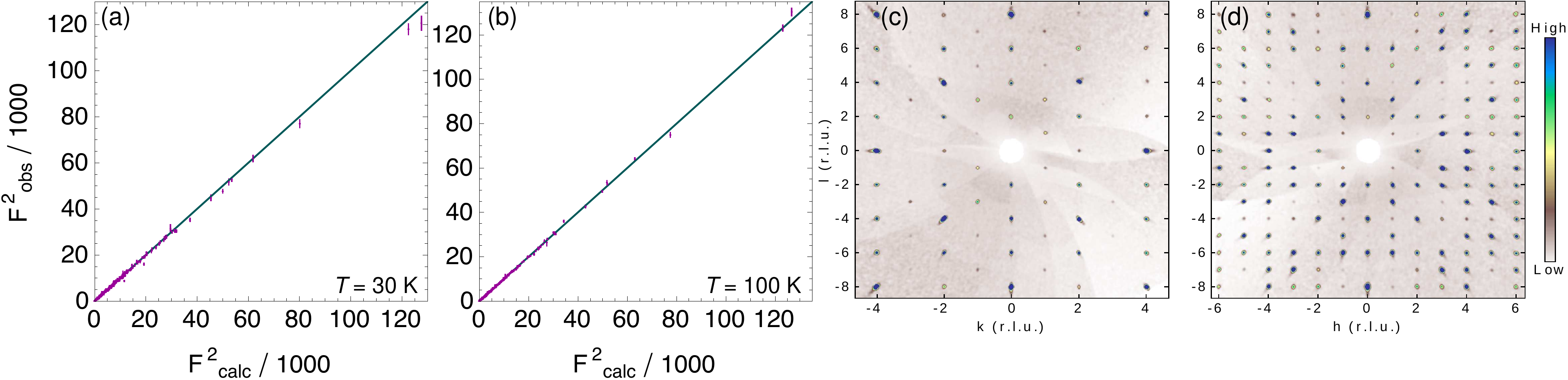}
  \caption{\label{fig:XRD}X-ray crystal structure refinements.  Comparison of the observed and calculated structure factors in our (a) 30-K and (b) 100-K single-crystal x-ray refinements, indicating excellent agreement.  Reciprocal space maps are shown of the (c) ($0kl$) and (d) ($h0l$) planes at 100\,K.}
\end{figure*}

To obtain the $g$-factor, high-field electron spin resonance (ESR) measurements were performed employing a transmission-type ESR spectrometer similar to that described in Ref.~\onlinecite{Zvyagin_INSR}, in magnetic fields up to 16\,T applied along the $a$ axis.  Measurements were done at a frequency of 356\,GHz, using a VDI microwave-chain radiation source (Virginia Diodes, Inc., USA).  An InSb hot-electron bolometer (QMC Instruments Ltd., UK) was used to record the spectra. The accuracy of the estimation of ESR fields and linewidths is better than $\pm 2$\%.

\section{Structure refinements\label{suppXRD}}

\begin{table}[b!]
  \caption{\label{tab:NPD}Details of the refinement of the crystal structure of antlerite based on our neutron powder diffraction data at 5.9\,K and single-crystal x-ray diffraction data at 30\,K.  These results are also described in {\tt Antlerite\_NPD\_5p9K.cif} (neutron) {\tt Antlerite\_XRD\_30K.cif} (x-ray).}
  \begin{tabular}{ll}\toprule\toprule
    Parameter & Value \\ \midrule
    Space group & $Pnma$ (No.~62)\\
    $a$ & 8.2097(5)\,\AA\\
    $b$ & 6.0337(5)\,\AA\\
    $c$ & 11.9527(10)\,\AA\\
    $V$ & 592.08(7)\,\AA$^3$\\
    $Z$ & 4\\
    Density & 4.0226(5)\,g/cm$^3$\\
    Reflections (neutron) & 181\\
    Reflections (x-ray) & 6795\\
    Unique reflections (x-ray) & 1070\\
    Unique reflections $I > 2\sigma$ (x-ray) & 874\\
    $hkl$ range probed (x-ray) & $-12\leq h\leq 11$,\\
    & $-7\leq k\leq 8$,\\
    & $-17\leq l\leq 17$\\
    $F(000)$ (x-ray) & 684\\
    $R$ (neutron) & 3.63\,\%\\
    $wR$ (neutron) & 4.64\,\%\\
    $R$ on reflections $I > 2\sigma$ (x-ray) & 2.56\,\%\\
    $wR$ on reflections $I > 2\sigma$ (x-ray) & 6.43\,\%\\ \bottomrule\bottomrule
  \end{tabular}
\end{table}

\begin{table}[b!]
  \caption{\label{tab:NPDatom}Refined atomic positions in synthetic antlerite based on our x-ray diffraction data at 30\,K and our neutron powder diffraction data at 5.9\,K, corresponding to the structure shown in Fig.~\ref{fig:fig1} in the main text.  Wy is the Wyckoff position. Cu, S, and O were refined with anisotropic thermal parameters, which are reported in Table~\ref{tab:U}; for these sites an equivalent $U_\text{equiv}$ is presented for $U$. The neutron sample was deuterated; the deuteration refined to 95.54(14)\,\%D and 4.46(14)\,\%H.}
  \begin{tabular}{l@{~\,}c@{~~}r@{.}lr@{.}lr@{.}lr@{.}l}\toprule\toprule
    Site & Wy & \multicolumn{2}{c}{$x$} & \multicolumn{2}{c}{$y$} & \multicolumn{2}{c}{$z$} & \multicolumn{2}{c}{$U$}\\ \midrule
    Cu1 & $4c$ & 0&00465(11) & 0&25 & 0&00134(8) & 0&0035(2)\\
    Cu2 & $8d$ & 0&29002(5) & 0&00302(10) & 0&12598(3) & 0&00353(17)\\
    S & $4c$ & 0&13130(16) & 0&25 & 0&36417(11) & 0&0034(5)\\
    O1 & $4c$ & 0&2631(5) & 0&25 & 0&2820(3) & 0&0057(18)\\
    O2 & $4c$ & 0&1997(5) & 0&25 & 0&4779(4) & 0&0061(18)\\
    O3 & $8d$ & 0&0319(3) & 0&0464(4) & 0&3482(2) & 0&0054(12)\\
    O4 & $4c$ & 0&2812(6) & 0&25 & 0&0250(4) & 0&0051(19)\\
    O5 & $4c$ & 0&7010(6) & 0&25 & 0&7779(4) & 0&0059(18)\\
    O6 & $8d$ & 0&0469(3) & 0&5064(6) & 0&1022(2) & 0&0048(10)\\
    D1 & $4c$ & 0&3588(9) & 0&25 & 0&9675(7) & 0&00582\\
    D2 & $4c$ & 0&2875(12) & 0&25 & 0&7699(7) & 0&02649\\
    D3 & $8d$ & 0&5121(8) & 0&0110(12) & 0&6721(4) & 0&02802\\ \bottomrule\bottomrule
  \end{tabular}
\end{table}

\begin{table*}[bthp]
  \caption{\label{tab:U}Refined anisotropic thermal parameters in synthetic antlerite based on our x-ray diffraction data at 30\,K.}
  \begin{tabular}{lr@{.}lr@{.}lr@{.}lr@{.}lr@{.}lr@{.}lr@{.}l}\toprule\toprule
    Site & \multicolumn{2}{c}{$U_{11}$} & \multicolumn{2}{c}{$U_{22}$} & \multicolumn{2}{c}{$U_{33}$} & \multicolumn{2}{c}{$U_{12}$} & \multicolumn{2}{c}{$U_{13}$} & \multicolumn{2}{c}{$U_{23}$}\\ \midrule
    Cu1 & 0&0044(2) & 0&0020(2) & 0&0040(2) & 0&0 & $-$0&00118(20) & 0&0\\
    Cu2 & 0&00323(16) & 0&00259(17) & 0&00479(17) & $-$0&0001(2) & $-$0&00089(15) & 0&0006(2)\\
    S & 0&0029(5) & 0&0035(5) & 0&0039(5) & 0&0 & 0&0002(5) & 0&0\\
    O1 & 0&0049(18) & 0&0060(18) & 0&0063(18) & 0&0 & 0&0032(15) & 0&0\\
    O2 & 0&0073(19) & 0&0059(18) & 0&0049(18) & 0&0 &$-$0&0007(16) & 0&0\\
    O3 & 0&0040(11) & 0&0034(13) & 0&0087(11) & $-$0&0008(9) & 0&0004(10) & $-$0&0001(9)\\
    O4 & 0&0052(18) & 0&0047(18) & 0&0054(19) & 0&0 & 0&0014(17) & 0&0\\
    O5 & 0&0065(19) & 0&0055(18) & 0&0056(18) & 0&0 & 0&0015(16) & 0&0\\
    O6 & 0&0045(10) & 0&0059(11) & 0&0040(10) & $-$0&0002(13) & $-$0&0002(9) & $-$0&0008(13)\\ \bottomrule\bottomrule
  \end{tabular}
\end{table*}

Key details of our refinement of the crystal structure of deuterated antlerite based on neutron powder diffraction data at 5.9\,K and single-crystal x-ray diffraction data at 30\,K are summarized in Table~\ref{tab:NPD}, and Crystallographic Information Files (CIF) are also provided as part of the Supplementary Material.  The refined atomic positions are reported in Table~\ref{tab:NPDatom}, and anisotropic thermal parameters in Table~\ref{tab:U}.  The deuteration refined to 95.54(14)\,\%D and 4.46(14)\,\%H.  Figure \ref{fig:XRD} shows selected results from these refinements.  Reciprocal space maps of the $(0kl)$ and $(h0l)$ planes at 100\,K, shown in Figs.~\ref{fig:XRD}(c) and \ref{fig:XRD}(d), respectively, show sharp spots, indicating good crystal quality.  Plots of the observed {\itshape vs}.\ calculated structure factors $F^2_\text{calc}$ and $F^2_\text{calc}$, shown in Figs.~\ref{fig:XRD}(a) and \ref{fig:XRD}(b) for 30 and 100\,K, respectively, indicate the excellent quality of the refinements.

\begin{table}[b!]
  \caption{\label{tab:NPD100}Details of the refinement of the crystal structure of antlerite based on our single-crystal x-ray diffraction data at 100\,K.  These results are also described in {\tt Antlerite\_XRD\_100K.cif}.}
  \begin{tabular}{ll}\toprule\toprule
    Parameter & Value\\ \midrule
    Space group & $Pnma$ (No.~62)\\
    $a$ & 8.2267(12)\,\AA\\
    $b$ & 6.0457(9)\,\AA\\
    $c$ & 11.9741(18)\,\AA\\
    $V$ & 595.55(15)\,\AA$^3$\\
    $Z$ & 4\\
    Density & 3.9562(10)\,g/cm$^3$\\
    Reflections & 15998\\
    Unique reflections & 1544\\
    Unique reflections $I > 2\sigma$ & 1210\\
    $hkl$ range probed (x-ray) & $-13\leq h\leq 13$,\\
    & $-10\leq k\leq 9$,\\
    & $-19\leq l\leq 19$\\
    $F(000)$ (x-ray) & 684\\
    $R$ on reflections $I > 2\sigma$ (x-ray) & 2.36\,\%\\
    $wR$ on reflections $I > 2\sigma$ (x-ray) & 4.45\,\%\\ \bottomrule\bottomrule
  \end{tabular}
\end{table}

\begin{table}[b!]
  \caption{\label{tab:NPDatom100}Refined atomic positions in synthetic antlerite based on our x-ray diffraction data at 100\,K.  Wy is the Wyckoff position. Cu, S, and some O sites were refined with anisotropic thermal parameters, which are reported in Table~\ref{tab:U100}; for these sites an equivalent $U_\text{equiv}$ is presented for $U$. As expected for an x-ray refinement, hydrogen positions are closer to their nearest oxygen atoms than the positions derived from neutron data at lower temperature.  Without neutron data, it was not possible to refine the H2 position.}
  \begin{tabular}{lcr@{.}lr@{.}lr@{.}lr@{.}l}\toprule\toprule
    Site & Wy & \multicolumn{2}{c}{$x$} & \multicolumn{2}{c}{$y$} & \multicolumn{2}{c}{$z$} & \multicolumn{2}{c}{$U$}\\ \midrule
    Cu1 & $4c$ & 0&00465(4) & 0&25 & 0&00133(3) & 0&00355(7)\\
    Cu2 & $8d$ & 0&29003(2) & 0&00300(4) & 0&12598(2) & 0&00362(6)\\
    S & $4c$ & 0&13126(7) & 0&25 & 0&36416(5) & 0&00329(10)\\
    O1 & $4c$ & 0&2635(2) & 0&25 & 0&28210(14) & 0&0060(3)\\
    O2 & $4c$ & 0&1994(2) & 0&25 & 0&47802(15) & 0&0060(3)\\
    O3 & $8d$ & 0&03205(15) & 0&0469(2) & 0&34816(11) & 0&0055(2)\\
    O4 & $4c$ & 0&2816(2) & 0&25 & 0&02488(15) & 0&0048(3)\\
    O5 & $4c$ & 0&7014(2) & 0&25 & 0&77753(14) & 0&0054(3)\\
    O6 & $8d$ & 0&04683(15) & 0&5064(2) & 0&10239(10) & 0&0049(2)\\
    H1/D1 & $4c$ & 0&355(6) & 0&25 & 0&981(4) & 0&037(13)\\
    H2/D2 & $4c$ & \multicolumn{2}{c}{---} & \multicolumn{2}{c}{---} & \multicolumn{2}{c}{---} & \multicolumn{2}{c}{---}\\
    H3/D3 & $8d$ & 0&504(4) & 0&016(6) & 0&662(4) & 0&051(12)\\ \bottomrule\bottomrule
  \end{tabular}
\end{table}

\begin{table*}[htbp]
  \caption{\label{tab:U100}Refined anisotropic thermal parameters in synthetic antlerite based on our x-ray diffraction data at 100\,K.}
  \begin{tabular}{lr@{.}lr@{.}lr@{.}lr@{.}lr@{.}lr@{.}lr@{.}l}\toprule\toprule
    Site & \multicolumn{2}{c}{$U_{11}$} & \multicolumn{2}{c}{$U_{22}$} & \multicolumn{2}{c}{$U_{33}$} & \multicolumn{2}{c}{$U_{12}$} & \multicolumn{2}{c}{$U_{13}$} & \multicolumn{2}{c}{$U_{23}$}\\ \midrule
    Cu1 & 0&00449(12) & 0&00212(12) & 0&00405(11) & 0&0 & $-$0&00115(9) & 0&0\\
    Cu2 & 0&00332(9) & 0&00275(9) & 0&00478(9) & 0&00062(7) & $-$0&00086(6) & $-$0&00012(8)\\
    S & 0&0027(2) & 0&0033(2) & 0&0039(2) & 0&0 & 0&00040(19) & 0&0\\
    O1 & 0&0049(8) & 0&0067(8) & 0&0063(7) & 0&0 & 0&0030(6) & 0&0\\
    O2 & 0&0084(8) & 0&0052(8) & 0&0044(7) & 0&0 &$-$0&0019(6) & 0&0\\
    O3 & 0&0045(5) & 0&0043(6) & 0&0077(5) & 0&0001(4) & 0&0005(4) & $-$0&0012(4)\\
    O4 & 0&0054(8) & 0&0037(8) & 0&0052(7) & 0&0 & 0&0005(6) & 0&0\\
    O5 & 0&0077(8) & 0&0034(8) & 0&0051(7) & 0&0 & 0&0001(6) & 0&0\\
    O6 & 0&0048(5) & 0&0059(5) & 0&0040(4) & $-$0&0004(5) & $-$0&0003(4) & 0&0000(5)\\ \bottomrule\bottomrule
  \end{tabular}
\end{table*}

Our x-ray refinement at 100\,K is summarized in Tables~\ref{tab:NPD100}, \ref{tab:NPDatom100}, and \ref{tab:U100}.  With the obvious exception of the hydrogen sites, all atomic positions are well within the uncertainty of the values refined at lower temperature, but they are more precise at 100\,K due to more comprehensive coverage of reciprocal space.  This gives us additional confidence in the reliability of the data at low temperature, and excludes significant structural changes between 100\,K and the magnetic transitions.  Without neutron data at a comparable temperature it was possible to refine the positions of H1 and H3, but not H2.  The sensitivity of x-rays to electrons, rather than the nucleus, shifts these positions closer to their nearest oxygen atom than in the neutron refinements, but they remain a plausible $\sim$0.8\,\AA\ away.  Hydrogen positions derived from neutron data are expected to more accurately represent the position of H$^+$ --- neutrons scatter off the nucleus, while x-rays scatter off charge density, which in the case of H is both extremely sparse and shifted toward the nearest anion.  Therefore, x-ray-derived H positions are typically shifted by $\sim$0.1\,\AA, consistent with our results.

\begin{figure}[b!]
  \includegraphics[width=0.5\columnwidth]{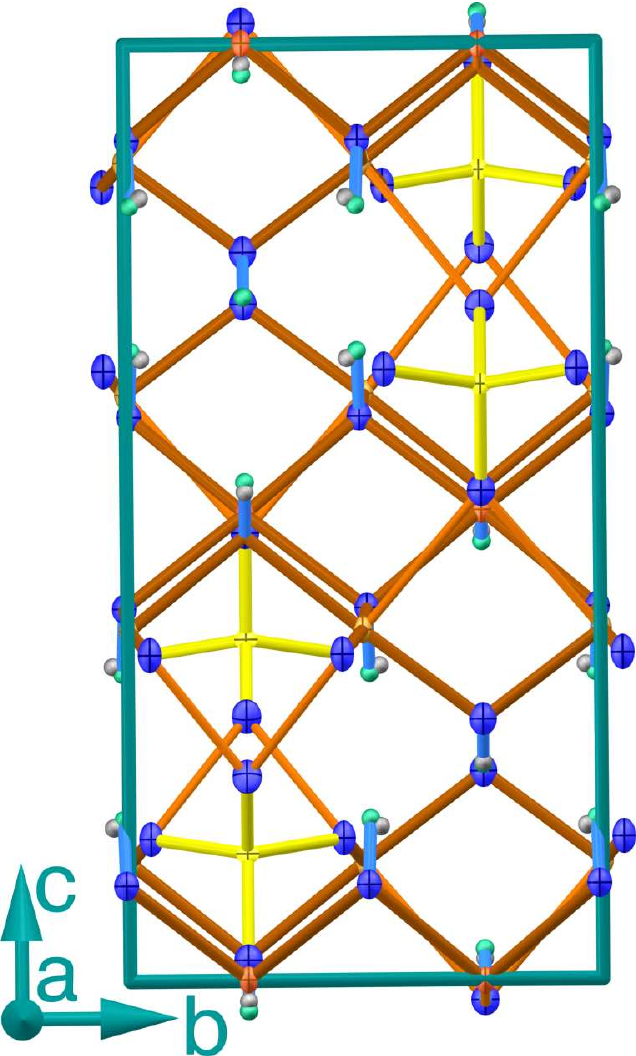}
  \caption{\label{fig:lattice}View of our refined crystal structure along the $a$ axis, where the hydrogen positions from Ref.~\cite{Vilminot2003} are shown in gray for comparison.}
\end{figure}

Our refined hydrogen positions are compared with those reported in Ref.~\onlinecite{Vilminot2003} in Fig.~\ref{fig:lattice}.  We find sites that are more symmetrically located relative to the host oxygen atoms.  This may be due to the higher temperature in the previous report.  The earlier data were collected at room temperature --- at lower temperatures the hydrogen atoms may sit closer to the bottom of their potential well, expected to be asymmetric, while refined atomic positions become significantly more precise on cooling as thermal motion freezes out.

\section{Magnetic Structure Refinement}

\begin{table}[htbp]
  \caption{\label{Mx}Refined magnetic moments in antlerite in the magnetic space group $Pn'm'a'$ (number 62.449) from data collected on the D1B diffractometer at 2.2\,K.}
  \begin{tabular}{lr@{.}l}\toprule\toprule
    Component & \multicolumn{2}{c}{Moment ($\mu_\text{B}$)}\\ \midrule
    Cu1 $M_x$ & $\pm 0$&54(17)\\
    Cu1 $M_y$ & 0&0\\
    Cu1 $M_z$ & $\pm 0$&59(14)\\
    Cu2 $M_x$ & 0&0\\
    Cu2 $M_y$ & $\pm 0$&38(5)\\
    Cu2 $M_z$ & $\pm 0$&89(9)\\
\bottomrule\bottomrule
  \end{tabular}
\end{table}

The magnetic irreducible representation was determined using {\sc SARAh}\,\cite{Sarah} to be $\Gamma_2$ and the propagation vector was identified using {\sc K\_Search}, part of the {\sc FullProf} Suite\,\cite{FullProf}.  The magnetic space group is $Pn'm'a'$ (number 62.449).  Refined magnetic moments for our low-temperature magnetic structure based on data collected at 2.2\,K at D1B are provided in Table~\ref{Mx}.  A magnetic CIF describing these results, {\tt Antlerite\_2p2K\_mag.mcif}, is provided as part of these Supplementary Materials.

\section{Electron Spin Resonance\label{suppESR}}

\begin{figure}[b!]
  \includegraphics[width=\columnwidth]{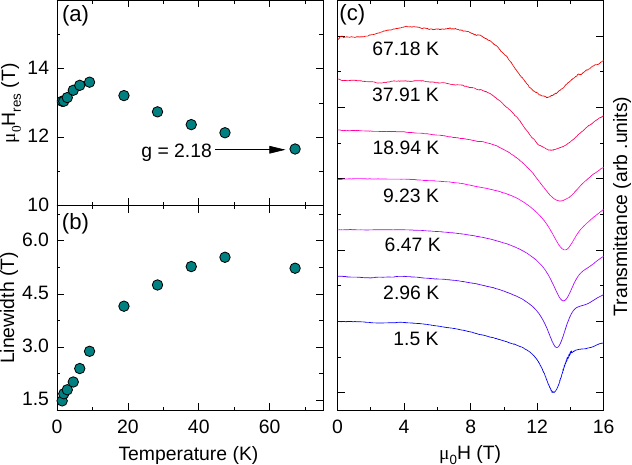}
  \caption{\label{fig:ESR}Temperature dependence of the (a) resonance field and (b) ESR linewidth, taken at a frequency of 356\,GHz with magnetic field applied along the $a$ axis. (c) Corresponding examples of ESR spectra.}
\end{figure}

Examples of ESR spectra are shown in Fig.~\ref{fig:ESR}(c).  The measurements revealed a relatively broad resonance line. The value $g = 2.18$ measured at a temperature of 70\,K [Fig.~\ref{fig:ESR}(a)] was used for calculation of the high-field magnetization.  With decreasing temperature, the ESR lineshape exhibits a significant narrowing [Fig.~\ref{fig:ESR}(b)], suggesting exchange interactions along the antiferromagnetic (AFM) chains as the main cause\,\cite{OA}.  We note that the temperature dependence of the linewidth is characteristic of antiferromagnetic $S=\frac12$ Heisenberg spin chains\,\cite{OA}.

\section{Higher-Energy Inelastic Neutron Scattering Data}

\begin{figure}[htbp]
  \includegraphics[width=\columnwidth]{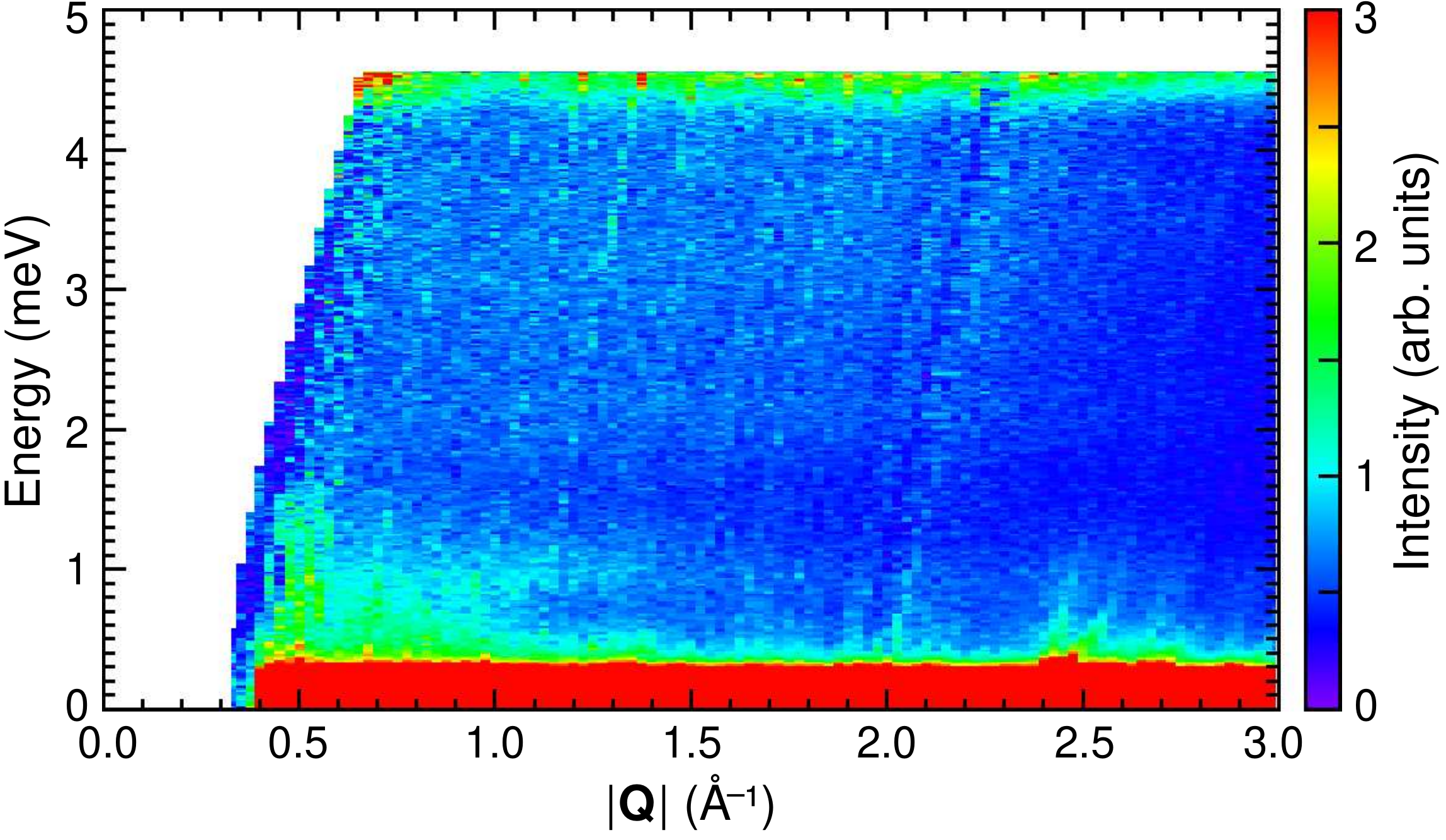}
  \caption{\label{moreFOCUS}Powder INS spectrum of deuterated antlerite at 1.5}
\end{figure}

Additional powder inelastic neutron scattering data collected on FOCUS at 1.5\,K using a neutron wavelength of 3.7\,\AA\ are shown in Fig.~\ref{moreFOCUS}.  The shorter wavelength allows access to higher energies.  No additional spin-wave features beyond those discussed in the main text are visible above 2\,meV.

\section{DFT calculations\label{suppDFT}}

\subsection{Band structure}

\begin{figure}[b!]
\includegraphics[width=\columnwidth]{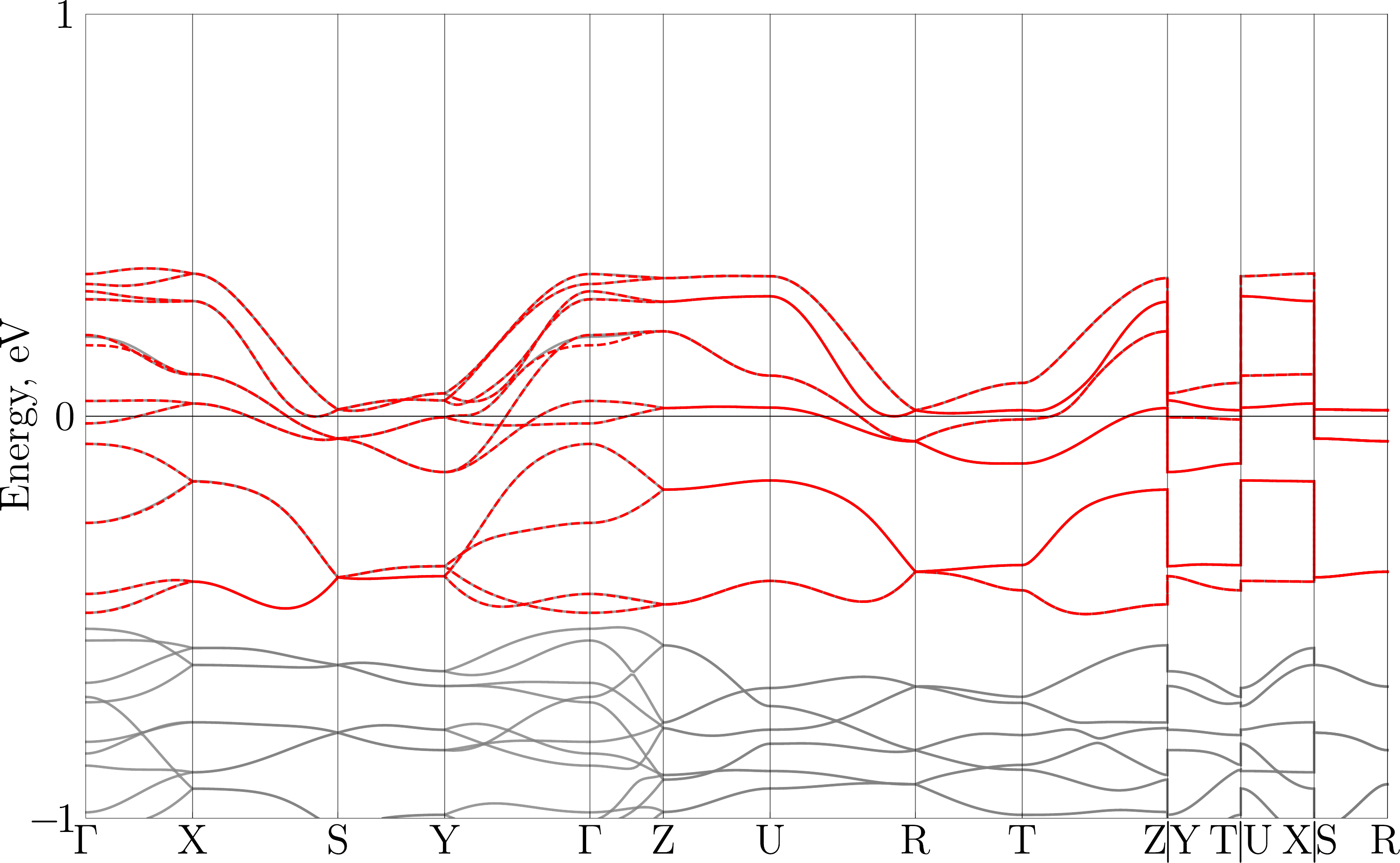}
\caption{The GGA band structure (gray) of antlerite and Fourier transform of the Wannier Hamiltonian (red). The Fermi level is at zero energy.\label{fig:bands}}
\end{figure}

The nonmagnetic band structure was calculated for the experimental crystal
structure of antlerite on a $k$-mesh of $9 \times 12 \times 6$ points.  The
underestimation of electronic correlations in GGA gives rise to a metallic
solution signalled by the bands which cross the Fermi level,
see~Fig.~\ref{fig:bands}. Antlerite is transparent and green, and there is no evidence for an electronic contribution to the specific heat, consistent with strongly insulating behavior, so we conclude that this metallic band structure is spurious, most likely due to strong electron correlations.  An inability to deal with strong correlations is a general shortcoming of band structure calculation techniques, which is commonly addressed by adding the Hubbard $U$ by hand\,\cite{Anisimov1991}.

By inspecting atomic and orbital characters of the
respective bands at the Fermi level, we find that they correspond to the antibonding combination
of Cu $d_{x^2 - y^2}$ and O $p_\sigma$ orbitals.  The $dp_\sigma$ hybridization
allows us to introduce an effective one-orbital model which captures the
low-energy physics of antlerite.  To this end, we constructed Cu-centered
Wannier functions, using $d_{x^2 - y^2}$ as the initial projector, following
the procedure described in Ref.~\cite{FPLO-WF}.  Consequently, we
obtain a Hamiltonian described in the basis of Wannier functions. A Fourier
transform of this Hamiltonian leads to excellent agreement with the GGA
bands, as seen in Fig.~\ref{fig:bands}.  The Wannier Hamiltonian contains hopping integrals that underpin our spin models, and the excellent quality of its band structure fit demonstrates that it captures the essential physics of the full GGA result.  

\subsection{Exchange couplings}	

\begin{figure}[b!]
\includegraphics[width=0.75\columnwidth]{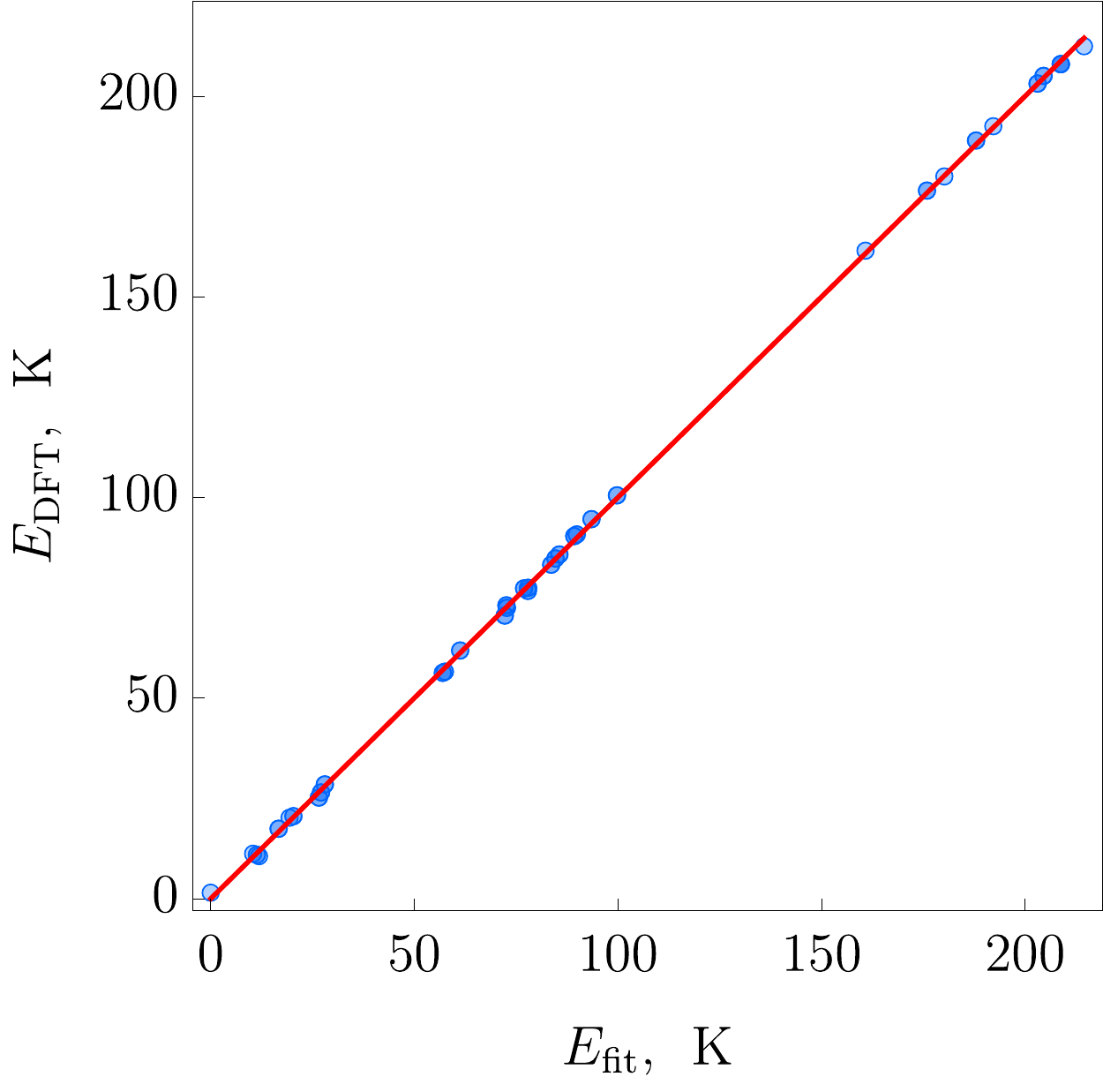}
\caption{Scatter plot of calculated ($E_\text{DFT}$) and fitted ($E_\text{fit}$) energies for 64 spin-polarized states for $U_d=7.5$\,eV.
Since the calculated and fitted energies both contain large nonmagnetic contributions, we subtract the minimal energy from both.\label{fig:fit_en}}
\end{figure}

	Exchange integrals were obtained by a least-squares solution of a redundant system of linear equations parameterized by GGA+$U$ total energies using Coulomb repulsion $U_d$ values of 7.5, 8.5, and 9.5\,eV. For each $U_d$, 64 magnetic configurations were considered. The fully localized limit was chosen for the double counting correction. The energies obtained from GGA+U calculations ($E_\text{DFT}$) and from fitting the model parameterized with exchange integrals ($E_\text{fit}$) is shown in Fig.~\ref{fig:fit_en}.  Results are summarized in Table~\ref{tab:Js}.  The Weiss temperature $\theta_{\text{W}}$ was calculated as a weighted sum of all exchanges normalized for $S\!=\!\frac12$; the weights equal the multiplicities of the respective exchanges. 
        
\begin{table}[t!]
  \caption{Values of magnetic exchanges and the Weiss temperature $\theta_{\text{W}}$ as a function of the Coulomb repulsion $U_d$ employed in our GGA+$U$ calculations. See main text for the notation of the magnetic exchanges. Values are in kelvins.\label{tab:Js}}
\begin{tabular}{rddd}\toprule\toprule
	$U_d$ & \multicolumn{1}{c}{7.5\,eV} & \multicolumn{1}{c}{8.5\,eV} & \multicolumn{1}{c}{9.5\,eV} \\ \midrule
	$J_1$ & -22.7    & -26.3    & -28.7 \\
	$J_2$ & -10.8    & -11.3    & -11.6 \\
	$J_3$ & 12.1     & 8.6      & 5.8   \\
	$J_4$ & 15.9     & 10.6     & 6.7   \\
	$J_5$ & 74.1     & 48.0     & 26.9  \\
	$J_6$ & 28.7     & 24.5     & 20.6  \\
	$J_6'$ & 7.3      & 6.1      & 5.1   \\ 
	$J_7$ & 1.3      & 1.1      & 1.0   \\
	$J_7'$ & 0.2      & 0.1      & 0.1   \\ \midrule
	$\theta_{\text{W}}$ & 23.3     & 14.2     & 7.1 \\ \bottomrule\bottomrule
\end{tabular}
\end{table}

\section{Ground-state calculations\label{suppDMRG}}

\subsection{Parameter dependence of the magnetic structure}

\begin{figure}[t!]
\includegraphics[width=\columnwidth]{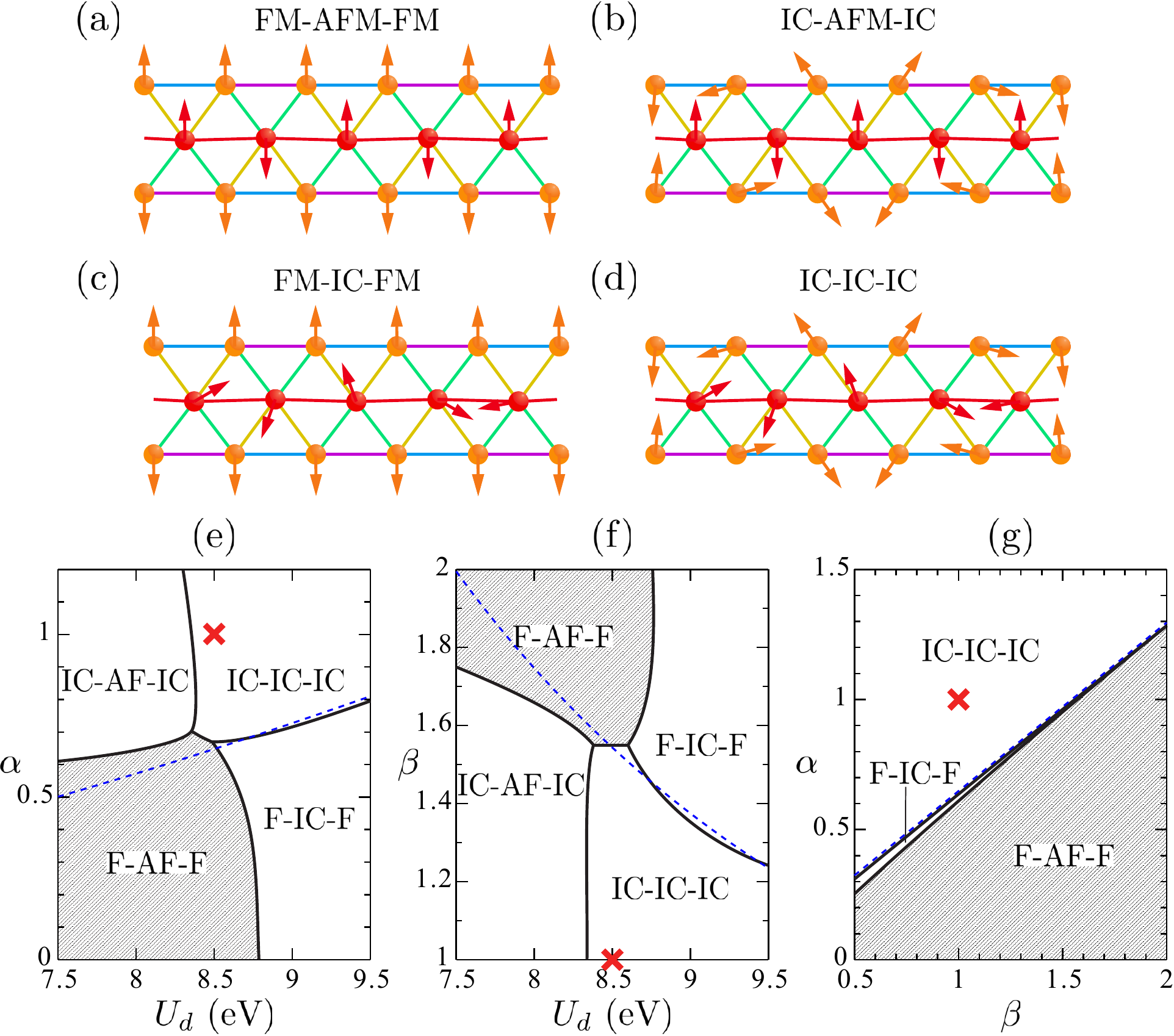}
\caption{(a-d) Schematic representations of possible magnetic ground states
	in the parameter region of \antH\
	deduced by DFT calculations.
	(e) $\alpha$--$U_d$ phase diagram with $\beta=1$.
	(f) $\beta$--$U_d$ phase diagram with $\alpha=1$.
	(g) $\alpha$--$\beta$ phase diagram. Red cross denotes
	the DFT parameter set for $U_d=8.5$\,eV. See also text.\label{fig:gs}}
\end{figure}

Antlerite, \antH, is a quasi-1D material, and its
magnetic properties may be well described by a three-leg
$S=1/2$ Heisenberg zigzag ladder modified by additional further-neighbor interactions. While the experimentally observed noncollinearity hints at the relevance of anisotropic exchanges, a numerical estimation of anisotropic terms requires noncollinear full-relativistic calculations that are computationally too expensive. We also note that such terms are typically too small to fundamentally change the nature of the ground state.  A full estimation of the anisotropic components would be an interesting avenue for future research. To study the parameter dependence of the magnetic
structure we calculated the static structure factor using the density-matrix
renormalization group method with a $40 \times 3$ open cluster. We
identified four possible magnetic phases in the realistic
parameter region for \antH\ deduced by the DFT calculations.
This is a consequence of a delicate balance of frustrated interactions
in this system. The four phases can be explained by a combination of two
possible states in each of the legs: the central leg exhibits either
AFM order or an incommensurate spiral (IC) state,
while the side legs are either ferromagnetic (FM) or in an IC state.
Schematic representations of the four states are shown in Fig.~\ref{fig:gs}(a-d).

Magnetic diffraction data indicate that \antH\ is
in the FM-AFM-FM state at low temperature, where we use the notation $X$-$Y$-$X$ to denote outer legs in state $X$ and the central leg in state $Y$.  The DFT
parameter set gives an IC-AFM-IC state for $U_d<8.35$\,eV and an IC-IC-IC state
for $U_d>8.35$\,eV. Nevertheless, as mentioned in the main text, the
experimental FM-AFM-FM ground state can be readily reproduced after a small
modification of the DFT parameters. For example, an IC
state on the side leg is driven, roughly speaking, by a competition between the second-neighbor AFM
interaction $J_6'$ and first-neighbor FM interactions $J_1$ and $J_2$.
One may naively expect that FM order of the side leg can be achieved by
reducing $J_6'$ or by increasing $J_1$, $J_2$. We thus investigated the
ground state as a function of $J_6'$, $J_1$, and $J_2$.
For convenience, the original $J_6'$ and $(J_1, J_2)$ are replaced by
$\alpha J_6'$ and $\beta (J_1, J_2)$, respectively, while the other DFT
parameters for a given $U_d$ are left unchanged.

Let us first consider adjusting only $J_6'$
by changing $\alpha$. The $U_d$--$\alpha$ phase diagram keeping
$J_1$ and $J_2$ unchanged, i.e., $\beta=1$, is shown in Fig.~\ref{fig:gs}(e).
We can see that FM order of the side legs is achieved for any plausible $U_d$ when
$\alpha$ is below $\sim0.6$. Similarly, as shown in
Fig.~\ref{fig:gs}(f), FM order is obtained on the side legs when $\beta$
is increased without unchanging $J_6'$ --- i.e., for $\alpha=1$. 
Starting with the DFT parameters at $U_d=8.5$\,eV, the experimental
FM-AFM-FM state is realized at $\alpha \lesssim 0.67$ in Fig.~\ref{fig:gs}(e) or at $\beta \gtrsim 1.55$ in Fig.~\ref{fig:gs}(f).
Although these modifications may seem large,
the experimental FM-AFM-FM state can be reproduced by, to cite one example,
only a $20\%$ reduction of $J_6'$ and a simultaneous $20\%$ enhancement of $J_1$ and $J_2$
from the DFT parameters ($U_d=8.5$\,eV). Furthermore, any reduction of $J_6$, enhancement
of $J_7$, or increase of $|J_3-J_4|$ would also act to 
stabilize the FM-AFM-FM state. It is possible to realize
the experimental FM-AFM-FM state with only several-$\%$ modifications
of the DFT parameters when all of the DFT parameters are tuned.

Additionally, it is instructive to consider the decoupled limit of the legs.
In this limit, the FM critical point of the side leg is exactly obtained
because the quantum fluctuations completely vanish at this point.
For a leg with alternating FM interactions $\beta J_1$, $\beta J_2$ and second-neighbor AFM interaction $\alpha J_6'$, the critical point is
obtained via~\cite{J1J1J2}
\begin{align}
 \frac{\alpha J_6'}{\beta J_1}=\frac{J_2}{2(J_1+J_2)}\text{.}
 \label{FMcritical}
\end{align}
This relation is plotted as dashed blue lines in Fig.~\ref{fig:gs}(e--g), where
we can see that it closely tracks the phase boundaries involving a
transition of the side legs from IC to FM order. This may
indicate that the decoupled-leg limit gives a good approximation for the FM
instability of the side legs. However, a more interesting finding is that a
direct transition from the IC-IC-IC to FM-AFM-FM states occurs around
$U_d=8.5$\,eV. In other words, the central leg is antiferromagnetically
aligned in concurrence with the FM transition of the side legs even though
the parameters $J_5$ and $J_6$ of the central leg are unchanged. 
This is related to a unique structure of the inter-leg interactions $J_3$
and $J_4$. More details are discussed in the next subsection.

It is also worth noting that the modified DMRG parameter set is near an instability to a Haldane phase. If $J_1/J_2$ or $J_3/J_4$ were a bit larger, adjacent spins on the $J_1$ bond would form a triplet pair, producing an effective spin-1 site. These spin-1 sites are coupled by AFM $J_6’$, so each side leg would be in the Haldane state rather than ferromagnetic. This is reflected in the particular pattern of spin correlations which manifests itself in the up-up-down-down state found by DFT based on the previously published structure. Alternatively, if the value of $J_6’$ is too small to stabilize the Haldane state, the system could be regarded as two coupled ferrimagnetic chains with alternating spins $S=1$ (side leg) and $S=\frac12$ (central leg) where each ferrimagnetic chain is formed along the $J_1$ and $J_4$ links. 

\subsection{Magnetization with magnetic field}

\begin{figure}[t!]
\includegraphics[width=\columnwidth]{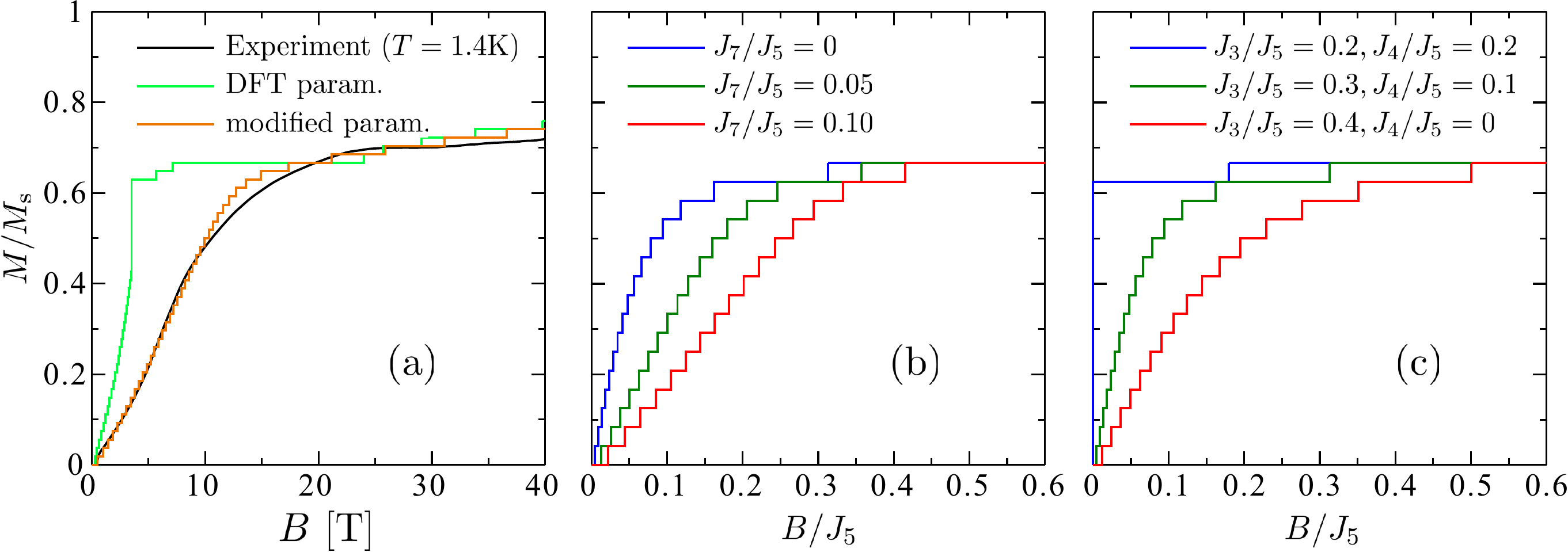}
\caption{(a) Experimental magnetization curve at 1.4\,K on \antH\ powder compared to those
	calculated by DMRG with the DFT parameter set ($U_d=8.5$\,eV) and
	a modified parameter set: $J_1=-25.2$, $J_2=-16.8$, $J_3=14.7$,
	$J_4=6.3$, $J_5=42.0$, $J_6=10.5$, $J_6'=1.7$, and $J_7=6.7$\,K.
        (b) $J_7$-dependence of the magnetization
	curve with fixed $J_1=J_2=-0.2J_5$, $J_3=0.3J_5$, $J_4=0.1J_5$, and
	$J_6=J_6'=0$. (c) $|J_3-J_4|$-dependence
	of the magnetization curve with fixed $J_1=J_2=-0.2J_5$,
        $J_6=J_6'=0$, $J_7=0$. \label{fig:mag}}
\end{figure}

Field-dependent magnetization measurements performed on antlerite powder up to 58\,T are shown in Fig.~\ref{fig:mag}(a) as well as Fig.~\ref{fig:mh} in the main text --- powder was chosen since the theoretical model is isotropic and would otherwise require powder averaging of the data.  The vertical scale here was set by fitting the experimental data to the DMRG results.
The magnetization of Cu$_3$SO$_4$(OH)$_4$ exhibits a peculiar field
dependence, the most notable feature of which is a broad plateau
at $M/M_{\rm s}=2/3$. This is essentially due to a large difference
in susceptibility between the central and side legs. With increasing field, 
the nearly FM side legs are polarized first, then the AFM central
leg begins to polarize at higher field. The location of this feature is well described by the DFT
parameter set. However, there are some quantitative discrepancies
compared to the experimental curve: the predicted magnetization increases too
rapidly at low fields and its slope where the side legs saturate
is too steep. Nevertheless, as mentioned in the main text, these
discrepancies can be resolved by a minor modification of the DFT
parameter set. The three magnetization curves are compared in
Fig.~\ref{fig:mag}(a). We here briefly explain why such a modification
was applied.

We first provide guidelines on how to obtain a slower increase of magnetization at low fields. The simplest way is to increase the AFM
interaction between the two side legs, i.e., $J_7$. Since each side leg is
in a FM state, their magnetization is controlled only by AFM inter-leg
interactions~\cite{Li2CuO2}. This effect is demonstrated in Fig.~\ref{fig:mag}(b).
We can see that the susceptibility, i.e., $dM/dB$, at $B=0$ is inversely proportional to $J_7$. Note that, for simplicity, we neglect the the second-neighbor interactions $J_6$, $J_6'$ and the alternation of
$J_1$ and $J_2$
because they are not essential in this analysis.
Another way to push the onset of the kink in magnetization to higher fields
is to increase the difference between $J_3$ and $J_4$.
In our spin model, each side spin is coupled to two central spins by
$J_3$ and $J_4$. The central leg consists of two sublattices
due to AFM order, and these two central spins belong to different sublattices.
Thus, for $J_3=J_4$, the inter-leg interactions $J_3$ and $J_4$ effectively
cancel out, and the magnetization curve
behaves as if there were no interactions between the central and side legs. 
This is confirmed by an abrupt jump in $M$ at $B=0$ in the case of
$J_3/J_5=J_4/J_5=0.2$ [Fig.~\ref{fig:mag}(c)]. Once the cancellation
balance is disrupted, i.e., $J_3 \ne J_4$, AFM fluctuations arise
between the central and side legs. As a result, the magnetization
increases more slowly with field for larger $|J_3-J_4|$ as shown in
Fig.~\ref{fig:mag}(c). Finally, we note that an increase of $J_6'$ could also lead to a slower
increase of the magnetization; however, $J_6'$ destroys the FM order of the side
legs, so an increase of $J_6'$ is not considered in detail.

\begin{figure}[b!]
  \includegraphics[width=0.8\columnwidth]{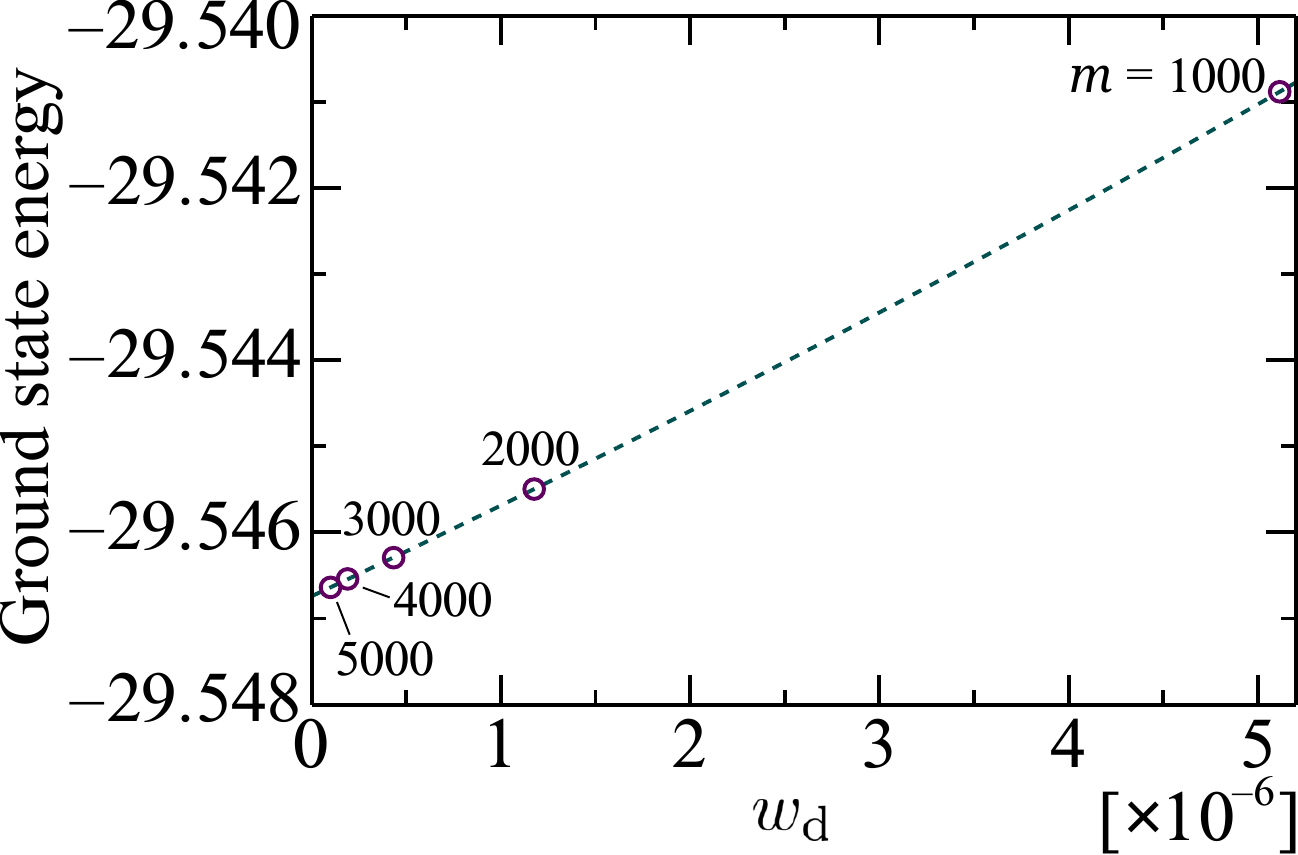}
  \caption{Ground state energy in the $S^z=0$ sector using a $36 \times 3$ periodic cluster as a function of discarded weight $w_{\rm d}$. The number of states $m$ retained in the renormalization procedure is also denoted. \label{fig:DMRG}}
\end{figure}

Next, we discuss the origin of the rounding of the magnetization curve around $B=15$\,T, which implies
an asymptotic saturation of the side legs. Note that the
effect of thermal fluctuations must be small because the measurement temperature $T=1.4$\,K is much smaller than the main interactions $J_1$ and
$J_2$ of the side legs. In general, such rounding of a magnetization curve
is only observed when significant spin anisotropy exists. In the case of antlerite, this
rare feature arises from the unique structure of the inter-leg interactions
$J_3$ and $J_4$ in our three-leg ladder. In Fig.~\ref{fig:mag}(c), we find that the rounding becomes more obvious with increasing
$|J_3-J_4|$. Let us now consider what happens. For the sake of clarity,
we assume translational-symmetry-broken AFM order on the central leg
and consider the case of $J_3/J_5=0.4$ and $J_4/J_5=0$. As the applied
field increases, the staggered magnetization on the central leg will rotate
to be more (anti-)parallel to the field.
With our maximally imbalanced $J_3$ and $J_4$, one side leg will be coupled only to up spins in the central leg,
while the other will be coupled only to down spins. These have different stability in high field, so
the susceptibility of one side leg will be suppressed and the other enhanced,
leading to asymptotic behavior of magnetization curve near the
saturation.

\subsection{Further details of the DMRG simulations}

We provide here further details on our DMRG simulations. One of the most difficult DMRG calculations in this paper is the estimation of the ground state energy in the $S^z=0$ sector using a $36 \times 3$ periodic cluster. This energy value is used to estimate the magnetization curve in Fig.~2(b) of the main text. A folding technique of the periodic cluster is applied to improve the accuracy of this calculation~\cite{Qin1995}. We keep $m=5000$ states in the renormalization procedure and the discarded weight is $w_{\rm d}=1.002\times 10^{-7}$. In Fig.~\ref{fig:DMRG} the ground state energy is plotted as a function of the discarded weight. We find that the convergence at $m\to\infty$ is sufficient with keeping $m=5000$ even in the most difficult case. For the calculations with an open cluster we keep up to $m=3000$ states and the discarded weight is less than $\sim 10^{-9}$.

\end{document}